\documentstyle[twocolumn,prb,aps,epsf]{revtex}

\begin{document}
\draft
\input{psfig}
\title{Phase Separation under Shear in Two-dimensional Binary Fluids}
\author{A.J. Wagner\footnotemark[2]
and J.M. Yeomans}
\address{Department of Physics, Theoretical Physics
1 Keble Rd. Oxford OX1 3NP, UK.
}
\date{\today}
\maketitle
\begin{abstract}
\renewcommand{\thefootnote}{\fnsymbol{footnote}}
\footnotetext[2]{MIT room 13-5157,77 Massachusetts av.,Cambridge MA 02139, U.S.A., e-mail: awagner@mit.edu}
\noindent
We use lattice Boltzmann simulations to study the effect of shear on
the phase ordering of a two-dimensional binary fluid. 
The shear is imposed by generalising the lattice Boltzmann algorithm
to include Lees-Edwards boundary conditions. We show how the interplay
between the ordering effects of the spinodal decomposition and the
disordering tendencies of the shear, which depends on the shear rate and the
fluid viscosity, can lead to a state of dynamic equilibrium where
domains are continually broken up and re-formed. 

\end{abstract}
~\\


\section{Introduction}

We present numerical results for the effect of shear flow on the
spinodal decomposition of a two-dimensional binary fluid using lattice
Boltzmann simulations.  We show how the lattice Boltzmann algorithm
can be generalised to allow the introduction of the Lees-Edwards
boundary conditions, which are commonly used in molecular dynamics
simulations to impose a shear flow without introducing walls.  Results
are presented showing how the competition between the ordering effects
of the free energy and the disordering effects of the shear influences the
spinodal decomposition and phase ordering of the fluid. 
For a recent review see Onuki \cite{onukiref}.

When a binary fluid consisting of an equal amount of two components, 
A and B say, is rapidly cooled below the critical temperature 
it phase separates into an A-rich and B-rich phase.  Once 
well-defined domains of each phase are formed the typical domain 
size grows according to a power law 
\begin{equation}
R(t) \sim t^\alpha 
\end{equation} 
where $\alpha$ is the growth exponent\cite{bray}.
$\alpha$ depends on the growth mechanism, which is dictated by the
surface tension, viscosity and diffusivity of the fluid, and the
time elapsed after the quench. In two-dimensional systems diffusive
Lifshitz-Slyozov growth gives $\alpha = \frac{1}{3}$ while
hydrodynamics can lead to faster growth with $\alpha = \frac{2}{3}$.

The most obvious effect of shear flow on the domain growth is that 
the growing domains are elongated in the direction of the flow, 
leading to an anisotropic morphology.  Experiments in three dimensions have 
shown that a string-like phase of thin domains oriented parallel to
the shear can be formed in strong shear\cite{hashimoto}.  Such domains, 
which would normally be expected to be unstable due to the 
Rayleigh instability, appear to be stabilised by the shear, although 
very recent experiments show that they can eventually break 
up in strong shear \cite{matsuzaka}.

This apparent stabilization suggests the possibility of a dynamic
equilibrium when stretching and breaking of the domains as the result
of the shear is balanced by their growth due to the thermodynamic
driving force and to the coalescence of the domains, which
can itself be driven by the shear.  This was first proposed by Ohta
and Nozaki \cite{otha} on the basis of two-dimensional simulations
using a cell dynamic approach.  These simulations, however, did not
include hydrodynamics.

Simulations of phase separation under shear which include
hydrodynamics are limited.  Rothman performed early work using lattice
gas cellular automata in two and three dimensions and was able to see
the anisotropy of the growth \cite{olson,rothman2}.  Wu {\it et. al.}
undertook Langevin simulations in two and three dimensions and report
the eventual formation of a string phase in three dimensions \cite{skrdla}.
Padilla and Toxvaerd performed molecular dynamics simulations on a
two-dimensional Lennard-Jones system, again pointing out the
anisotropic nature of the domain growth \cite{padilla}.  In the simulations a
peak was seen in the excess shear viscosity as a function of time
corresponding to the increase in the lengths of interfaces in the
system.  However, there seems to be no evidence for a shear-induced
dynamic equilibrium.

Here we simulate phase separation under shear using a lattice
Boltzmann scheme in the same spirit as the model introduced by
Orlandini {\it et. al.}, which imposes phase separation by defining
the fluid equilibrium as the minimum of an input free energy
\cite{enzo,swift2}.  This method has been very successful in obtaining
results for phase separation in the absence of shear \cite{a_scaling}.
A particular advantage of the approach is that the fluid viscosity and
diffusivity can be tuned, and this has allowed us to compare
simulations for parameter values where diffusive or hydrodynamic phase
separation dominates.  We find either phases striped in the shear
direction, or a dynamic equilibrium where the length scales remain
approximately constant in time, depending on the relative strengths of
the shear and the ordering.

The lattice Boltzmann approach is described in \S 2.  Because 
this is a lattice rather than particulate simulation 
method, it is not immediately obvious how to define Lees-Edwards 
shear boundary conditions.  An approach for doing this is given
in \S 3.  In particular it is necessary to generalize the 
normal definition of the lattice Boltzmann equilibrium distribution.
In \S 4 we define suitable measures to characterise the 
anisotropic morphology of the spinodal decomposition patterns 
when shear is applied.  The results of our simulations are 
contained in \S 5, where the effect of shear is compared for
different fluid viscosities.  \S 6 summarises the 
results and discusses outstanding questions.

\section{The Lattice Boltzmann Approach}

The starting point for lattice Boltzmann simulations\cite{doolen} 
is the evolution
equation, discrete in space and time, for a set of distribution
functions, ${f_i}$, each associated with a velocity vector, ${\bf v}_i$. For
the sake of simplicity we consider a single relaxation time, the 
so-called BGK approximation
\cite{quian}. The evolution equation for the
$\{f_i\}$ is 
\begin{equation}
f_i({\bf x}+ {\bf v}_i \Delta t, t+\Delta t) -f_i({\bf x},t)=
\frac{\Delta t}{\tau_1} (f_i^0-f_i),
\label{LBE1}
\end{equation}
where ${\bf x}$ is a lattice point, $\Delta t$ is the time step, and
${\bf v_i} \Delta t$ is normally constrained to be a lattice
vector. The relaxation time is $\tau_1$ and $f_i^0$ is the
equilibrium distribution. For a two-component system a second,
equivalent equation is also needed
\begin{equation}
g_i({\bf x}+ {\bf v}_i \Delta t, t+\Delta t) -g_i({\bf x},t)=
\frac{\Delta t}{\tau_2} (g_i^0-g_i).
\label{LBE2}
\end{equation}
Physical quantities are defined as moments of the distribution
functions. To model the isothermal flow of a binary mixture of
components A and B, we choose
\begin{equation}
\sum_i f_i = n,\;\;\;\;\;
\sum_i f_i {\bf v}_i = n {\bf u},\;\;\;\;\;
\sum_i g_i = \varphi,
\label{eqn3}
\end{equation}
where $n$ is the total density field, ${\bf u}$ is the velocity field
and $\varphi$ is the field corresponding to the difference in the
density of components A and B. 

We require mass conservation for both components and momentum
conservation for the bulk. This is equivalent to 
constraining the equilibrium distributions to obey
\begin{equation}
\sum_i f_i^0 = n, \;\;\;\;\;
\sum_i g_i^0 = \varphi,\;\;\;\;\;
\sum_i f_i^0 {\bf v}_i = n {\bf u}.
\label{conserve_nu}
\end{equation}

We also need to define higher-order moments of the equilibrium
densities. The choice for these moments is within the free energy
lattice Boltzmann scheme used here\cite{enzo,swift2}
\begin{eqnarray}
\sum_i f_i^0 v_{i\alpha} v_{i\beta} &=& P_{\alpha\beta} + nu_\alpha
u_\beta, \label{cons1}\\
\sum_i g_i^0 v_{i\alpha} &=& \varphi u_\alpha,\\
\sum_i g_i^0 v_{i\alpha} v_{i\beta} &=& \Gamma \mu \delta_{\alpha\beta} 
+ \varphi u_\alpha u_\beta, \label{cons3}
\end{eqnarray}
where $P_{\alpha \beta}$ is the pressure tensor, $\Gamma$ is a mobility
parameter, $\mu$ is the chemical potential for the density difference
and $\delta$ is the Kronecker delta. The physical motivation for these
constraints is twofold; firstly to ensure the correct form of the
macroscopic equations of motion and secondly to reproduce the correct
thermodynamics of the 
binary mixture in equilibrium as discussed in more detail below.

Taylor-expanding the evolution equations (\ref{LBE1}) and (\ref{LBE2})
to second order in the derivatives gives the macroscopic equations of
motion for the binary fluid\cite{thesis}. These are the continuity 
equation for the
total density
\begin{equation}
\partial_t n+\partial_\alpha  n u_\alpha=0,
\label{cont_n}
\end{equation}
a convection-diffusion equation governing the evolution of the density
difference
\begin{equation}
\partial_t \varphi+\partial_\alpha  (\varphi u_\alpha) =
\omega_2
\left( \Gamma \nabla^2 \mu- \partial_\beta \left(
\frac{\varphi}{n} \partial_\alpha P_{\alpha\beta} \right)\right),
\label{cdeq}
\end{equation}
and, in the incompressible limit,
the incompressible Navier Stokes equations for a non-ideal
system
\begin{equation}
n \partial_t u_\alpha+n u_\beta \partial_\beta u_\alpha = 
-\partial_\beta P_{\alpha\beta} 
+ \frac{n \omega_1}{3}
\nabla^2 u_\alpha
+ O(\partial^3)
\label{NS}
\end{equation}
where $\omega_{1,2}=\tau_{1,2}-\Delta t/2$ and
the viscosity is given by $\nu=n\omega_1/3$.

The thermodynamic fields entering the simulation are the
pressure tensor and the chemical potential which follow from the free
energy of the system.
We consider the free energy of a simple binary fluid. 
A--A and B--B interactions are zero, but there is an A--B
repulsion $\lambda n_A n_B$ where $n_A$ and $n_B$ are the number
densities of A- and B-particles, respectively, and $\lambda$ is a
parameter describing the interaction strength. This system can be
described by the Landau free energy functional
\begin{equation}
\Psi = \int d{\bf r} \left\{ \psi(\varphi,n,T)
+ \frac{\kappa}{2} (\nabla \varphi)^2\right\} \label{totf}
\end{equation}
where $T$ is the temperature
and $\kappa$ is a measure of the excess interface free energy (surface
tension). The free energy density of the homogeneous system is
\cite{reichel}
\begin{eqnarray}
\psi(\varphi,n,T)&=&\frac{\lambda
n}{4}\left(1-\frac{\varphi^2}{n^2}\right) -Tn \nonumber\\
&&+\frac{T}{2}(n+\varphi) \ln\left(\frac{n+\varphi}{2}\right) \nonumber\\
&&+\frac{T}{2}(n-\varphi) \ln\left(\frac{n-\varphi}{2}\right). \label{bulkf}
\end{eqnarray}
For temperatures greater than a critical temperature $T_c=\lambda /2$
the system remains in a single phase. For
$T < T_c$ there is phase separation into two states with
$\varphi = \pm \varphi_0$.

From the free energy (\ref{totf}) we derive the local chemical
potential $\mu$ as the functional derivative of the total free energy
$\Psi$ with respect to the concentration difference field $\varphi({\bf
x})$
\begin{equation}
\mu({\bf x}) = \frac{\delta \Psi}{\delta \varphi({\bf x})}
= -\frac{\lambda}{2} \frac{\varphi}{n} + \frac{T}{2} \ln\left(
\frac{n+\varphi}{n-\varphi} \right) - \kappa \nabla^2 \varphi.
\end{equation}
Equilibrium corresponds to $\mu(\phi,n,T)=0$.

The derivation of the pressure tensor is slightly more involved and
is discussed in Appendix A\cite{yang}. We obtain
\begin{eqnarray}
P_{\alpha\beta}&=&
(n \partial_n \psi+ \varphi \partial_\varphi \psi -\psi)
\delta_{\alpha\beta} \nonumber\\
&&+ \kappa ( \partial_\alpha \varphi
\partial_\beta \varphi - \frac{1}{2} \partial_\gamma \varphi
\partial_\gamma \varphi \delta_{\alpha\beta} - \varphi \partial_\gamma
\partial_\gamma \varphi) \nonumber\\
&=&
(nT + \varphi \mu^0(x))\delta_{\alpha\beta}\nonumber\\
&&+ \kappa ( \partial_\alpha \varphi
\partial_\beta \varphi - \frac{1}{2} \partial_\gamma \varphi
\partial_\gamma \varphi \delta_{\alpha\beta} - \varphi \partial_\gamma
\partial_\gamma \varphi)\nonumber\\
&=& (nT + \varphi \mu(x)) \delta_{\alpha\beta}\nonumber\\
&&+\kappa (\partial_\alpha
\varphi \partial_\beta \varphi-\frac{1}{2} \partial_\gamma \varphi
\partial_\gamma \varphi \delta_{\alpha\beta})
\label{e:pressuretensor}
\end{eqnarray}
where the first term is the ideal gas pressure, the second term is the
osmotic pressure with $\mu^0=\partial_\phi \psi$ and the third term is
related to the surface tension. The osmotic
pressure was omitted in the original definition of the model
\cite{enzo,swift2}. The chemical potential and pressure tensor are
input to the lattice Boltzmann scheme through equations (\ref{cons1}) 
and (\ref{cons3}).
In equilibrium the simulated fluid minimises the free energy (\ref{totf}).

It remains only to define the equilibrium distributions $f_i^0$ and
$g_i^0$ introduced in the evolution equations (\ref{LBE1}) and
(\ref{LBE2}). Normally an
expansion to second order in the velocities is sufficient to reproduce
the constraints (\ref{conserve_nu}) -- (\ref{cons3})\cite{doolen}.
However, this ceases to be the
case when Lees-Edwards shear boundary conditions are introduced. In
the next section we discuss how the equilibrium distribution can be
defined to allow the use of Lees-Edwards boundary conditions.

\section{Shear boundary conditions} 

Possibly the easiest way to introduce shear flow in a lattice Boltzmann
simulation is to include walls moving in a lattice direction. Even for
a wall with neutral wetting, however, phase separation is
strongly enhanced at the walls and the wall effects easily dominate
the phase separation process for all but the largest systems. The
effect of walls on phase separation is an interesting phenomenon
in its own right, but it is not the process we are interested in
studying here.

The problem caused by explicit walls can be overcome in a relatively
simple and efficient manner by introducing a Klein-bottle symmetry to
the lattice. This is done by forcing the fluid to have a given
velocity along one line in the direction of the shear flow.  In a
one-component mixture this induces a linear velocity profile. For a
two-component mixture, however, the dynamics are influenced by the
V-shaped velocity profile at the forcing line because of the non-local
interactions. We used this algorithm to produce preliminary results
but it has no advantages over the method derived below.

A more regular shear flow can be produced by extending the idea of
Lees-Edwards boundary conditions, widely used in Molecular Dynamics
\cite{lee_edw}, to lattice Boltzmann simulations. Briefly, Lees and
Edwards simulated shear boundary conditions for a shear in the
$x$-direction in a simulation box of dimensions $(X,Y)$ by
introducing periodic boundary conditions in the
$y$-direction. Particles that left the box at the lower boundary for
position $(x,y=0)$ reappeared at the upper boundary at position $(x+ut
(\bmod X),y=Y)$ with a velocity that was changed by $v \rightarrow
v+u$.

To implement this idea for lattice Boltzmann simulations we are faced
with two difficulties. Firstly the densities are defined on a lattice
and the Lees-Edwards boundary conditions lead to densities defined between the
lattice points.  Secondly we need to impose a Galilean transformation
for the densities which are streamed across the lattice.

The non-fitting of the lattice is relevant for both the streaming and
for the calculation of derivatives at $y=1$ and $y=Y$. We solve
this problem by a linear interpolation scheme. For any density we
define
\begin{eqnarray}
f[x,y=0]&=&(1-R(ut)) f[x+I(ut),y=Y]\nonumber\\
&& + R(ut) f[x+I(ut)+1,y=Y]
\end{eqnarray}
where $I(z)$ is the largest integer with $I(z)<z$ and
$R(z)=z-I(z)$. If we pass the break in the lattice from the other side
we define similarly
\begin{eqnarray}
f[x,y=Y+1]&=&(1-R(ut)) f[x-I(ut),y=1]\nonumber\\
&& + R(ut) f[x-I(ut)-1,y=1].
\end{eqnarray}
These formulae are used both for the streaming of the
Galilean-transformed Boltzmann densities, $f_i$, and for the calculation
of density gradients.

\begin{figure}
\centerline{\psfig{figure=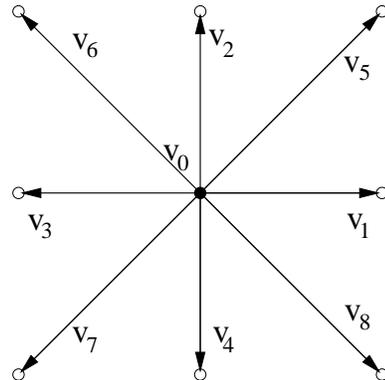,width=5cm}}
\caption{Our numbering of the velocity vectors in a nine-velocity model.}  
\label{fig:ninevel}
\end{figure}

It is rather more difficult to see how the Galilean
transformation should be defined. Let us consider the special case of
a two-dimensional, 
nine-velocity model where the velocities are numbered as indicated
in Figure \ref{fig:ninevel}. We need to perform a Galilean transformation on
the $\{5,2,6\}$ and the $\{7,4,8\}$ velocities as these
will carry mass and momentum across the boundaries. To define
the transformation we demand mass and $y$-momentum conservation
\begin{equation}
n_p \equiv f_5+f_2+f_6=f_5'+f_2'+f_6'
\end{equation}
an appropriate change in the $x$-momentum
\begin{equation}
(f_5-f_6)-(f'_5-f'_6)=n_p u
\label{udef}
\end{equation}
and conservation of the local pressure
\begin{eqnarray}
 P^p_{xx}&=&\sum{\frac{f_i}{{n_p}^2} (n{v_i}_x-n_p u_x)^2}\nonumber\\
&=&\frac{1}{{n_p}^2}\left((f_5(n_p-n_p u_x)^2+f_2(n_p
u_x)^2\right. \nonumber\\
&&\left. + f_6(-n_p-n_p u_x)^2\right)\nonumber\\
&=&\frac{1}{{n_p}^2}\left(f_5'(n_p-n_p u_x')^2+f_2'(n_p u_x')^2\right.
\nonumber\\
&&\left. + f_6'(-n_p-n_pu_x')^2 \right)
\end{eqnarray}
where the prime denotes the transformed quantities.

This system of equations can be solved to give a unique solution for
the Galilean-transformed densities $f_i'$
\begin{eqnarray}
f_2'&=& f_2 + 2 (f_5-f_6) u - n_p u^2,\label{cond1}\\
f_5'&=& f_5 + (-\frac{3}{2} f_5-\frac{1}{2}f_2+\frac{1}{2}f_6) u + 
\frac{n_p}{2} u^2,\\
f_6'&=& f_6 + (-\frac{1}{2}f_5 + \frac{1}{2} f_2+
\frac{3}{2} f_6)u + \frac{n_p}{2} u^2.\label{cond3}
\end{eqnarray} 
This definition can be extended to a Galilean transformation
for all densities and, equivalently, to a transformation in different lattice
directions.

In order for this transformation to make sense we need to make sure
that equation (\ref{udef}) is consistent with the definition of the
equilibrium distribution, $f_i^0$ in Eqn.~(\ref{LBE1}) {\it i.e.} 
that an equilibrium distribution
for a velocity $u$ Galilean transformed by a velocity $\Delta u$ is
equal to the equilibrium distribution for velocity $u+\Delta u$. It is
conventional to define the equilibrium distribution as a polynomial in
second order in u. A generic expansion is
\begin{eqnarray}
f_i^0&=&A_\sigma n+B_\sigma n u_\alpha {v_i}_\alpha + C_\sigma n u^2 
\nonumber\\
&&+
D_\sigma n u_\alpha u_\beta {v_i}_\alpha {v_i}_\beta+ G_{\sigma\alpha\beta}
n {v_i}_\alpha {v_i}_\beta
\label{equil}
\end{eqnarray}
where $A_\sigma,B_\sigma,C_\sigma,D_\sigma,G_{\sigma\alpha\beta}$ are
constants that have the absolute value of the corresponding velocity
vector $\sigma=|{\bf v}_i|$ as an index.  However, substituting
(\ref{equil}) into (\ref{udef}) shows that this equation is not
satisfied in equilibrium.  In practice this leads to a step in the
$u_x$ profile at the boundary.

There is, however, no {\it a priori} reason to use a second-order
expansion in the velocity for the equilibrium distribution. All that
is needed for a valid equilibrium distribution is that
(\ref{conserve_nu})--(\ref{cons3}) hold and that the distribution
obeys the conditions (\ref{cond1})--(\ref{cond3}).

Let $T_{\alpha\beta}=P_{\alpha\beta}/n$. Then, if we require,
\begin{equation}
\frac{f^0_1-f^0_3}{f^0_1+f^0_0+f^0_3}=u_x,\;\;\;\;\;\;\;\;\;\;\;
\frac{f^0_2-f^0_4}{f^0_2+f^0_0+f^0_4}=u_y
\label{ccond}
\end{equation}
\begin{eqnarray}
f^0_5+f^0_6-(f^0_5+f^0_6+f^0_2) (T_{xx}+u_x u_x)=0,\\
f^0_5+f^0_8-(f^0_5+f^0_8+f^0_1) (T_{yy}+u_y u_y)=0,\\
f^0_8+f^0_7-(f^0_8+f^0_7+f^0_4) (T_{xx}+u_x u_x)=0,\\
f^0_6+f^0_7-(f^0_6+f^0_7+f^0_3) (T_{yy}+u_y u_y)=0. \label{clast}
\end{eqnarray}
(\ref{ccond})-(\ref{clast}), together with (\ref{conserve_nu})--(\ref{cons3}) 
are a completely determined set of equations
with the solution
\begin{eqnarray}
f^0_0&=&n (1- T_{xx} -u_x^2)(1-T_{yy}-u_y^2),\nonumber\\
f^0_1&=&\frac{1}{2} n ( T_{xx}+u_x +u_x^2)(1-T_{yy}-u_y^2),\nonumber\\
f^0_2&=&\frac{1}{2} n ( T_{yy}+u_y +u_y^2)(1-T_{xx}-u_x^2),\nonumber\\
f^0_3&=&\frac{1}{2} n ( T_{xx}-u_x +u_x^2)(1-T_{yy}-u_y^2),\nonumber\\
f^0_4&=&\frac{1}{2} n ( T_{yy}-u_y +u_y^2)(1-T_{xx}-u_x^2),\nonumber\\
f^0_5&=&\frac{1}{4} n ( T_{xy} + T_{xx} T_{yy} +T_{yy} (u_x+u_x^2)
\nonumber\\&& +
	T_{xx} (u_y+u_y^2)+ u_x u_y (1+u_x+u_y+u_x u_y)),\nonumber\\
f^0_6&=&\frac{1}{4} n (-T_{xy} + T_{xx} T_{yy} +T_{yy} (-u_x+u_x^2)
\nonumber\\&& +
	T_{xx} (u_y+u_y^2)- u_x u_y (1-u_x+u_y-u_x u_y)),\nonumber\\
f^0_7&=&\frac{1}{4} n ( T_{xy} + T_{xx} T_{yy} +T_{yy} (-u_x+u_x^2)
\nonumber\\&& +
	T_{xx} (-u_y+u_y^2)+ u_x u_y (1-u_x-u_y+u_x u_y)),\nonumber\\
f^0_8&=&\frac{1}{4} n (-T_{xy} + T_{xx} T_{yy} +T_{yy} (u_x+u_x^2)
\nonumber\\&& +
	T_{xx} (-u_y+u_y^2)- u_x u_y (1+u_x-u_y-u_x u_y)).\nonumber
\end{eqnarray}
For this equilibrium distribution
\begin{equation}
\frac{f^0_5-f^0_6}{f^0_5+f^0_2+f^0_6}=u_x+\frac{T_{xy}}{T_{yy}+u_y+u_y^2}
\end{equation}
which is consistent with the Galilean transformation (\ref{udef}). 

For a two-component system we similarly define the $g_i^0$ using
\begin{equation}
\frac{g^0_1-g^0_3}{g^0_1+g^0_0+g^0_3}=u_x,\;\;\;\;\;\;\;\;\;
\frac{g^0_2-g^0_4}{g^0_2+g^0_0+g^0_4}=u_y
\label{cg1}
\end{equation}
and imposing
\begin{equation}
g_0 =\varphi- \ell \Gamma \mu -\varphi (u_x^2+u_y^2)\label{cge}
\end{equation}
where $\ell$ is a free parameter that can be used to improve stability
(we choose $\ell=1$). Solving equations (\ref{cg1}) and (\ref{cge})
and (\ref{conserve_nu})--(\ref{cons3}) gives
\begin{eqnarray}
g^0_1 &=&1/2 \{(\ell-1-u_x)\Gamma\mu+(1+u_x-u_y^2)\varphi u_x\},\nonumber\\
g^0_2 &=&1/2 \{(\ell-1-u_y)\Gamma\mu+(1+u_y-u_x^2)\varphi u_y\},\nonumber\\
g^0_3 &=&1/2 \{(\ell-1+u_x)\Gamma\mu-(1-u_x-u_y^2)\varphi u_x\},\nonumber\\
g^0_4 &=&1/2 \{(\ell-1+u_y)\Gamma\mu-(1-u_y-u_x^2)\varphi u_y\},\nonumber\\
g^0_5 &=&1/4 \{(2-\ell+u_x+u_y)\Gamma\mu+(1+u_x+u_y) \varphi u_x u_y\},\nonumber\\
g^0_6 &=&1/4 \{(2-\ell-u_x+u_y)\Gamma\mu-(1-u_x+u_y) \varphi u_x u_y\},\nonumber\\
g^0_7 &=&1/4 \{(2-\ell-u_x-u_y)\Gamma\mu+(1-u_x-u_y) \varphi u_x u_y\},\nonumber\\
g^0_8 &=&1/4 \{(2-\ell+u_x-u_y)\Gamma\mu-(1+u_x-u_y) \varphi u_x u_y\}.\nonumber
\end{eqnarray}
The macroscopic flow equations are unaffected by the choice of the further constraints
(\ref{ccond})--(\ref{clast}) and (\ref{cg1})--(\ref{cge}) or by
the detailed structure of the equilibrium distributions. Therefore,
these alterations in the model can change the numerical stability and the
behaviour of quantities like the spurious velocities, but they
leave the evolution of the macroscopic quantities unaffected, at least
to second order in the derivatives. 

\section{Measures for non-isotropic patterns}

To characterise the features of phase separation under shear it is
necessary to construct measures for  the length scales of the
sheared systems which will in general be anisotropic. 
Measures that are based on Fourier
transforms cannot be easily used for sheared systems because the
system is no longer periodic. 

Length scales derived from derivatives do not require periodicity.
Derivatives need to be evaluated for the algorithm and are
readily available. We define a tensor 
\begin{equation}
d_{\alpha \beta}
= \frac{\sum_{\bf x} 	\partial_\alpha^D \varphi({\bf x},t) 
			\partial_\beta^D  \varphi({\bf x},t)}{
	\sum_{\bf x} \varphi^2({\bf x},t)}\label{eqn:d}
\end{equation}
where $\partial_\alpha^D$ is the symmetric discrete derivative in
direction $\alpha$. Because the tensor is symmetric it can be
diagonalised to give two eigenvalues $\lambda_1,\lambda_2$ and an
angle $\theta^\star$
\begin{eqnarray}
\lambda_1 &=& \frac{d_{xx}+d_{yy}}{2} + 
\sqrt{\frac{(d_{xx}-d_{yy})^2}{4}+d_{xy}^2},\\
\lambda_2 &=& \frac{d_{xx}+d_{yy}}{2} -
\sqrt{\frac{(d_{xx}-d_{yy})^2}{4}+d_{xy}^2},\\
\theta^\star  &=& \tan^{-1}\left(\frac{d_{yy}}{d_{xy}-\lambda_2}\right).
\label{tetd}
\end{eqnarray}
The two eigenvalues give two orthogonal length
scales 
\begin{equation}
R^{\star}_1(t)=\frac{1}{\lambda_1(t)L_w},\;\;\;\;\;\;\;\; \label{rd1}
R^{\star}_2(t)=\frac{1}{\lambda_2(t)L_w}, 
\end{equation}
where $L_w$ is the interface width. It appears because
$d_{\alpha\beta}$ scales inversely with the interface
width\cite{thesis}.  
$L_w$, used as a constant here, could in principle be anisotropic. That this
anisotropy is not a strong effect can be seen by comparing these
length scales with scales that are explicitly independent of the
interface width.

One such measure is related to the lengths of the
interfaces in the system. The interface can be represented
by a set of contours. These contours consist of small line
segments $\vec{l}_i$ and the length of the interface can be written
\begin{equation}
L_I=\sum_i |\vec{l}_i|.
\end{equation}
In order to extract the preferred direction of the interface we define
the vector
\begin{equation}
\vec{D} = R^{-1}\left(\sum_i R(\vec{l}_i)\right).
\end{equation}
The operator $R$ is defined by
\begin{equation}
R(\vec{x})=|\vec{x}| \left(
\begin{array}{c}
\cos(2 \theta)\\
\sin(2\theta)
\end{array}\right)
\end{equation}
where $\theta$ is the angle between
the argument of R  and the $x$-axis.

$\vec{D}$ is a vector that is zero for isotropic closed contours
and which points in the average direction of the interface for
non-isotropic closed contours. Two length scales and an angle
that correspond to the intuitive result for
oriented rectangular objects can be defined from these measures
\begin{equation}
R^{\circ}_1=\frac{XY}{L_I+|\vec{D}|} \label{rc1},\;\;\;\;\;\;\;\;
R^{\circ}_2=\frac{XY}{L_I-|\vec{D}|},
\end{equation}
\begin{equation}
\theta^{\circ}  = 
\cos^{-1}\left(\frac{\hat{\vec{x}}.\vec{D}}{|\vec{D}|}\right).
\label{tetl}
\end{equation}
Thus we have defined two independent sets of measures for the structure of
non-isotropic patterns that will now be used to examine  spinodal
decomposition under shear.

\section{Simulation results}

For all the simulations we used a total density $n=2$, an
interaction parameter $\lambda=1.1$, which corresponds to a critical
temperature $T_c=0.55$, and a temperature $T=0.5$. The equilibrium
values of the order parameter were then $\varphi_0=\pm 1$. The mobility was
$\Gamma=2$, the relaxation time for the order parameter in
Eqn.~(\ref{LBE2})
was $\tau_2=1$ and the interface free energy parameter was
$\kappa=0.002$, which corresponds to an interface width of
approximately three lattice spacings. The relaxation parameter for the
total density Eqn.~(\ref{LBE1}), $\tau_1$, was varied: $\tau_1=100$
gave a high viscosity and $\tau_1=1$ an intermediate viscosity.

The shear transformation, S, is defined as 
\begin{equation}
S \left( \begin{array}{c}x \\y \end{array} \right) =
\left( \begin{array}{r}x+\dot{\gamma} t y \\y \end{array} \right).
\end{equation}
Shear flow applied to a system undergoing spinodal decomposition
stretches the original pattern. This effect is only relevant once the
deformation caused by the flow is of the same order or larger than the
deformation caused by the coarsening process. This requires
\begin{equation}
\dot{\gamma} t > 1 \label{sh1}.
\end{equation}
We therefore expect to observe the effect of the shear flow
for $t>1/\dot{\gamma}$.

\begin{figure}
\begin{center}
\begin{minipage}[t]{3cm}
\centerline{\psfig{figure=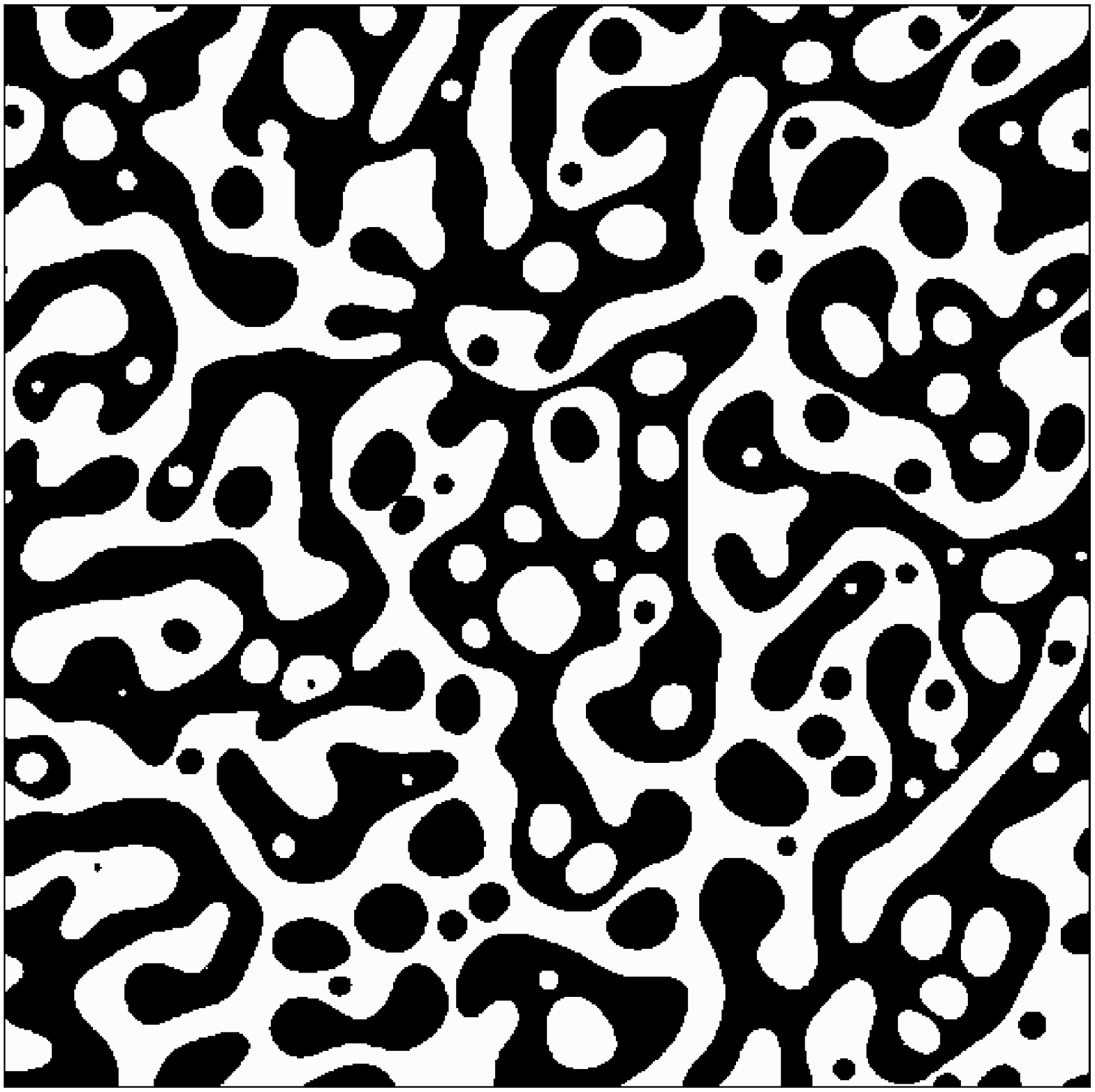,width=3cm}}
\begin{center} time=0 \end{center}
\end{minipage} \
\begin{minipage}[t]{3cm}
\centerline{\psfig{figure=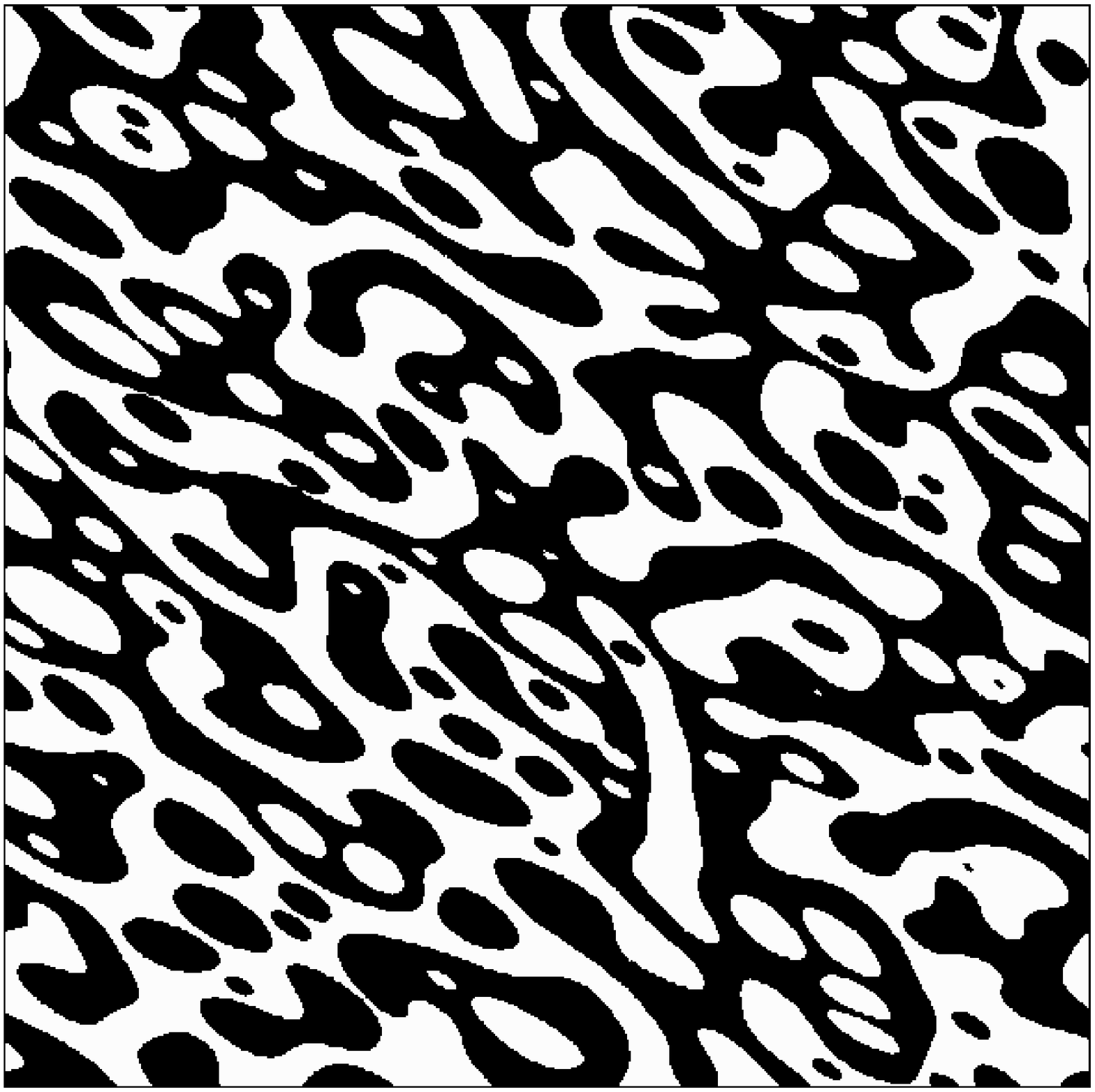,width=3cm}}
\begin{center} time=1 \end{center}
\end{minipage} \
\begin{minipage}[t]{3cm}
\centerline{\psfig{figure=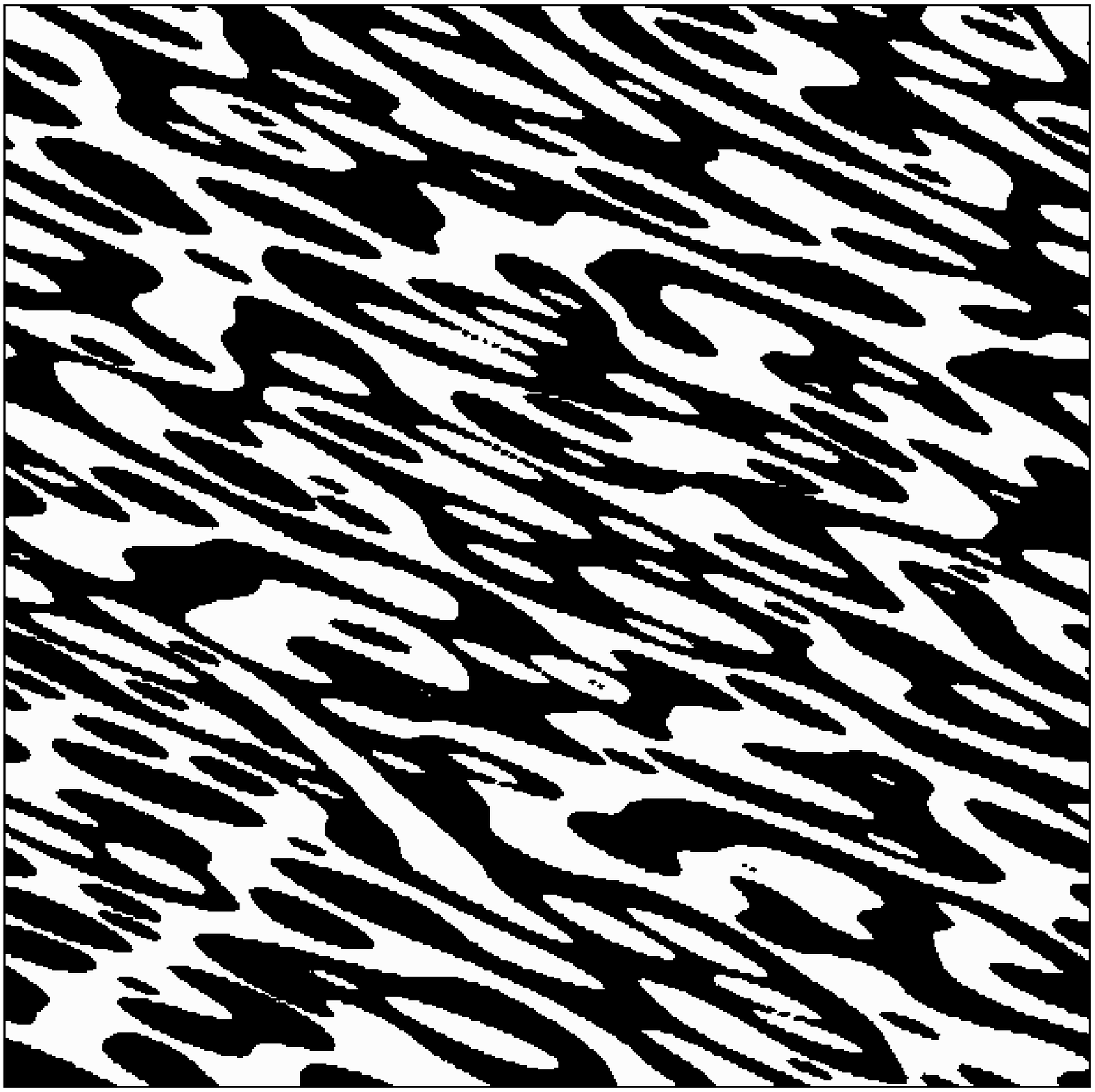,width=3cm}}
\begin{center} time=2 \end{center}
\end{minipage} \
\begin{minipage}[t]{3cm}
\centerline{\psfig{figure=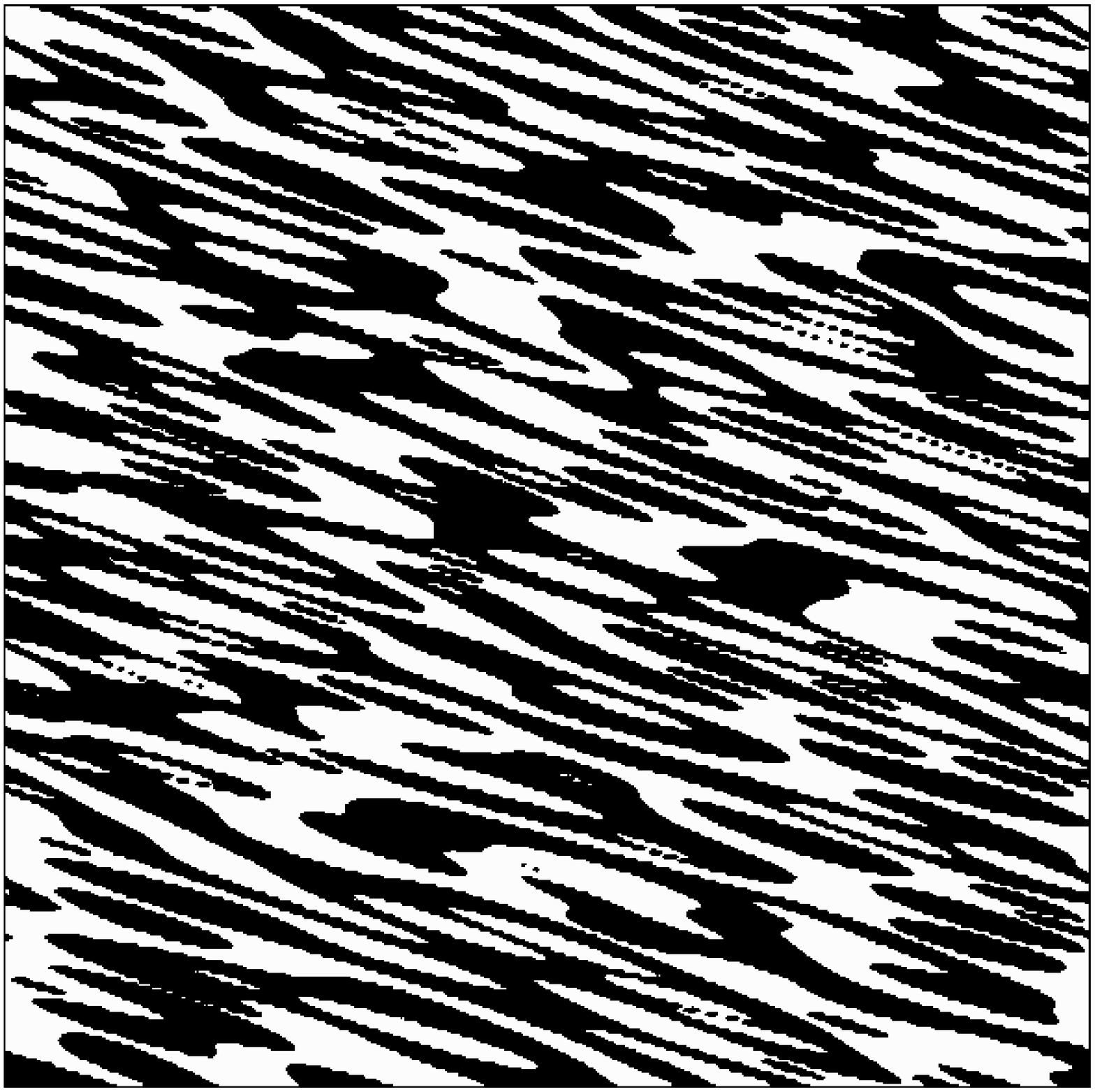,width=3cm}}
\begin{center} time=3 \end{center}
\end{minipage} \\
\begin{minipage}[t]{3cm}
\centerline{\psfig{figure=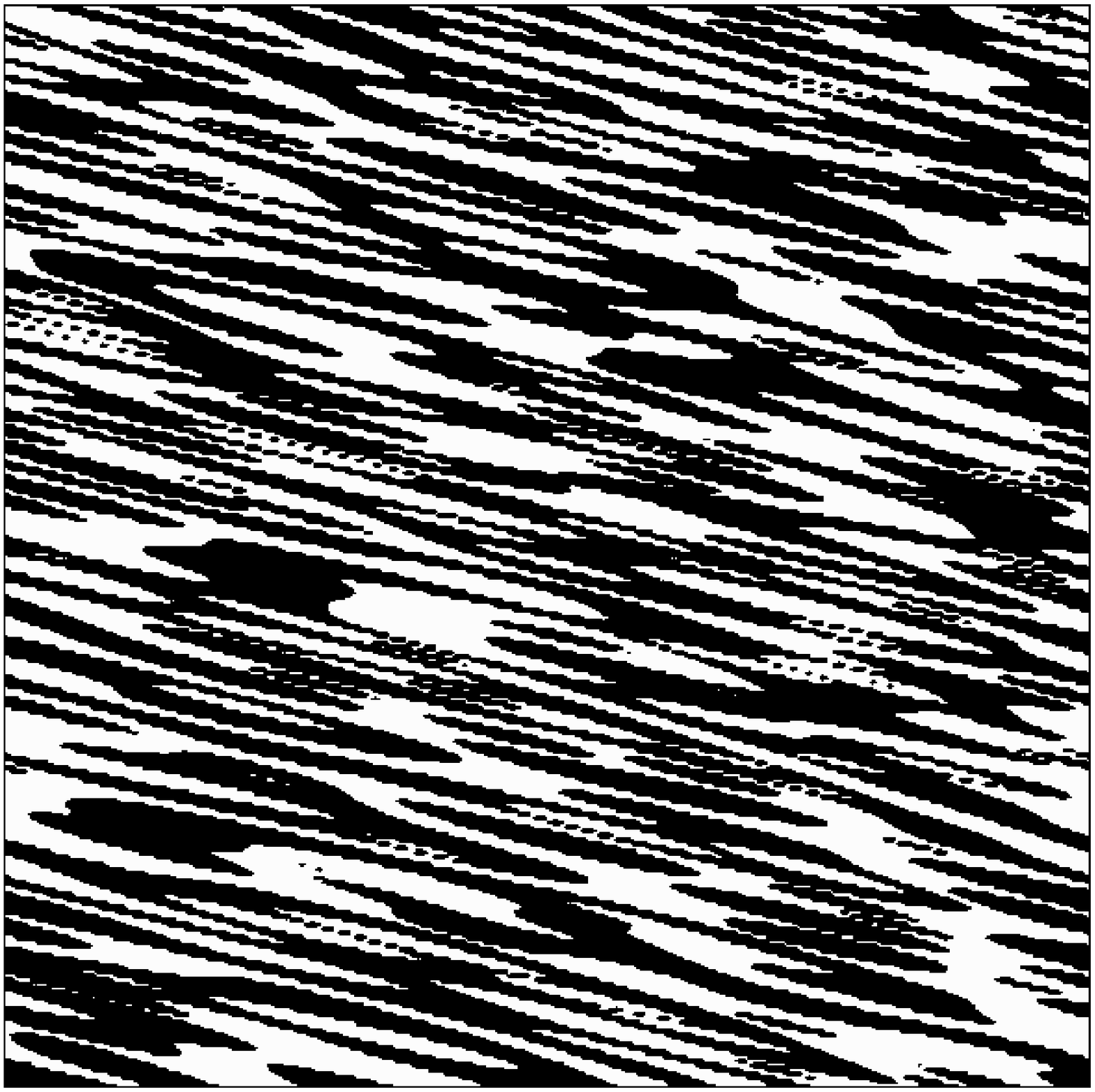,width=3cm}}
\begin{center} time=4 \end{center}
\end{minipage} \
\begin{minipage}[t]{3cm}
\centerline{\psfig{figure=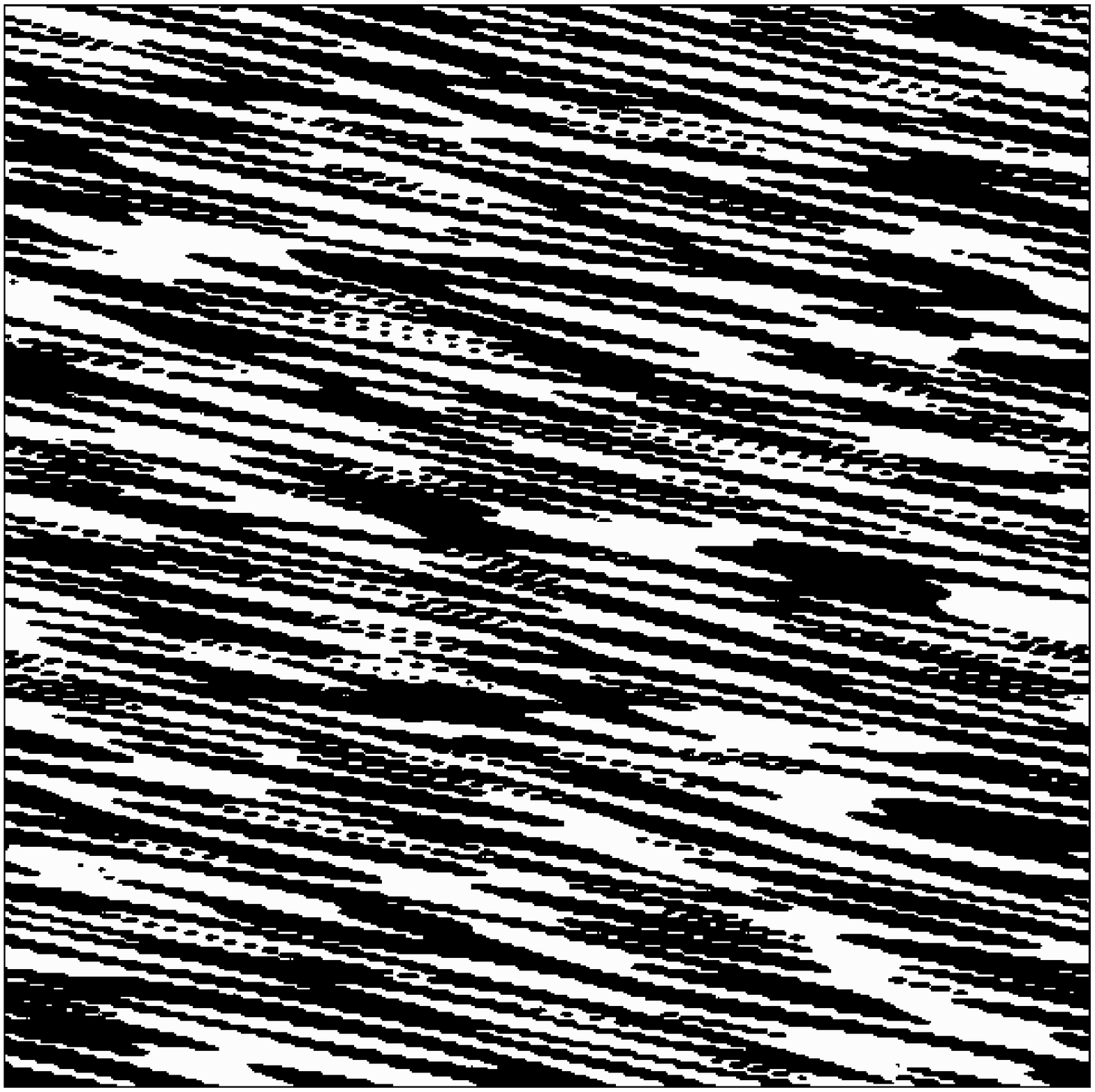,width=3cm}}
\begin{center} time=5 \end{center}
\end{minipage} \
\begin{minipage}[t]{3cm}
\centerline{\psfig{figure=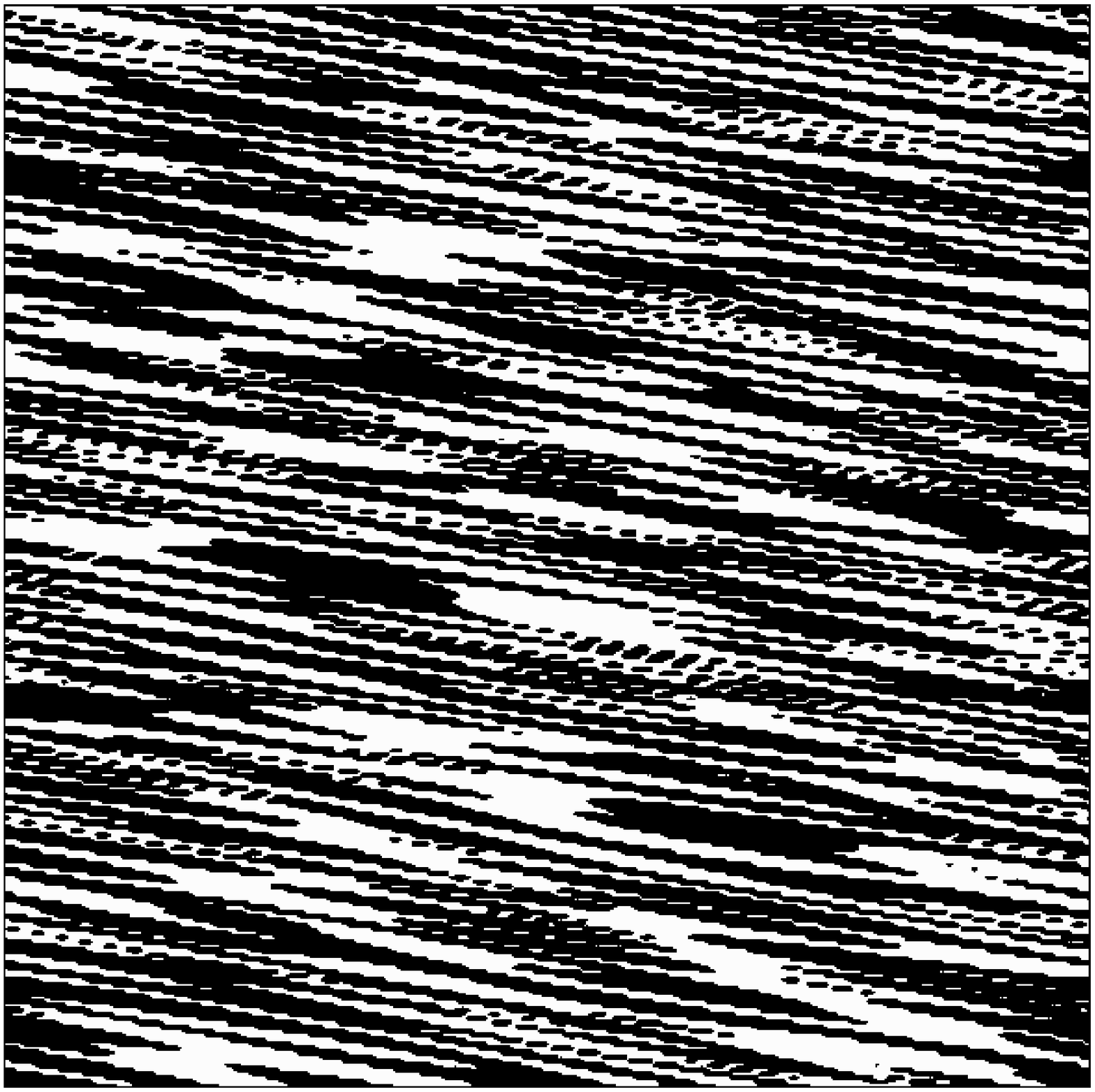,width=3cm}}
\begin{center} time=6 \end{center}
\end{minipage} \
\begin{minipage}[t]{3cm}
\centerline{\psfig{figure=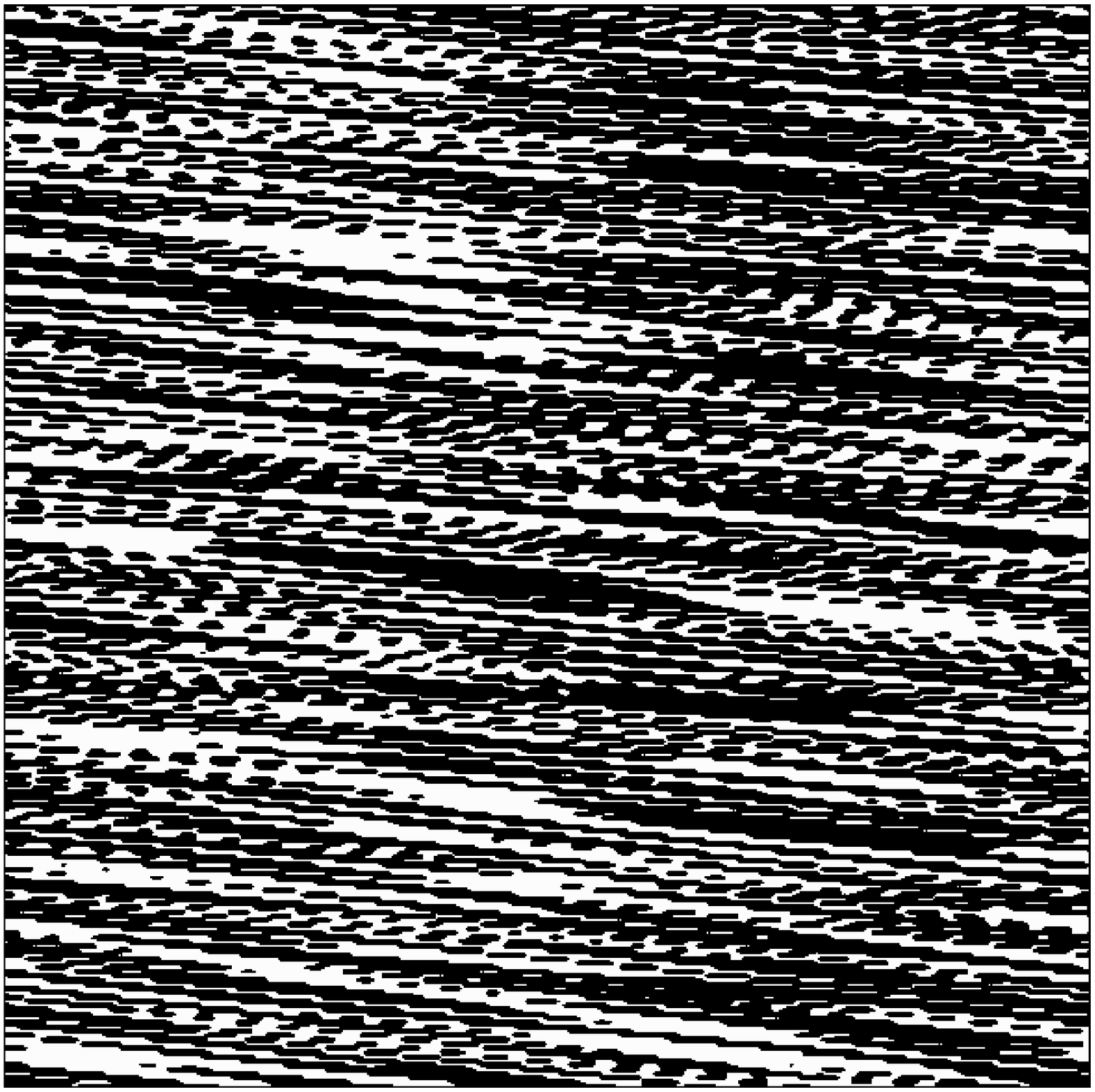,width=3cm}}
\begin{center} time=10 \end{center}
\end{minipage}
\end{center}
\caption{Applying shear (with shear rate $\dot{\gamma}=1$) to a system
without internal dynamics leads to homogenization.}
\label{fig:shear}
\end{figure}

To help understand the effect of shear-flow on a phase-separating
system let us first consider a pattern without any internal dynamics
that undergoes a shear transformation. This
transformation is illustrated in Figure \ref{fig:shear}, where we
start from a frozen spinodal decomposition pattern and show successive
iterations of a shear transformation with $\dot{\gamma}=1$.

\begin{figure}
\begin{minipage}{8cm}
  \centerline{\psfig{figure=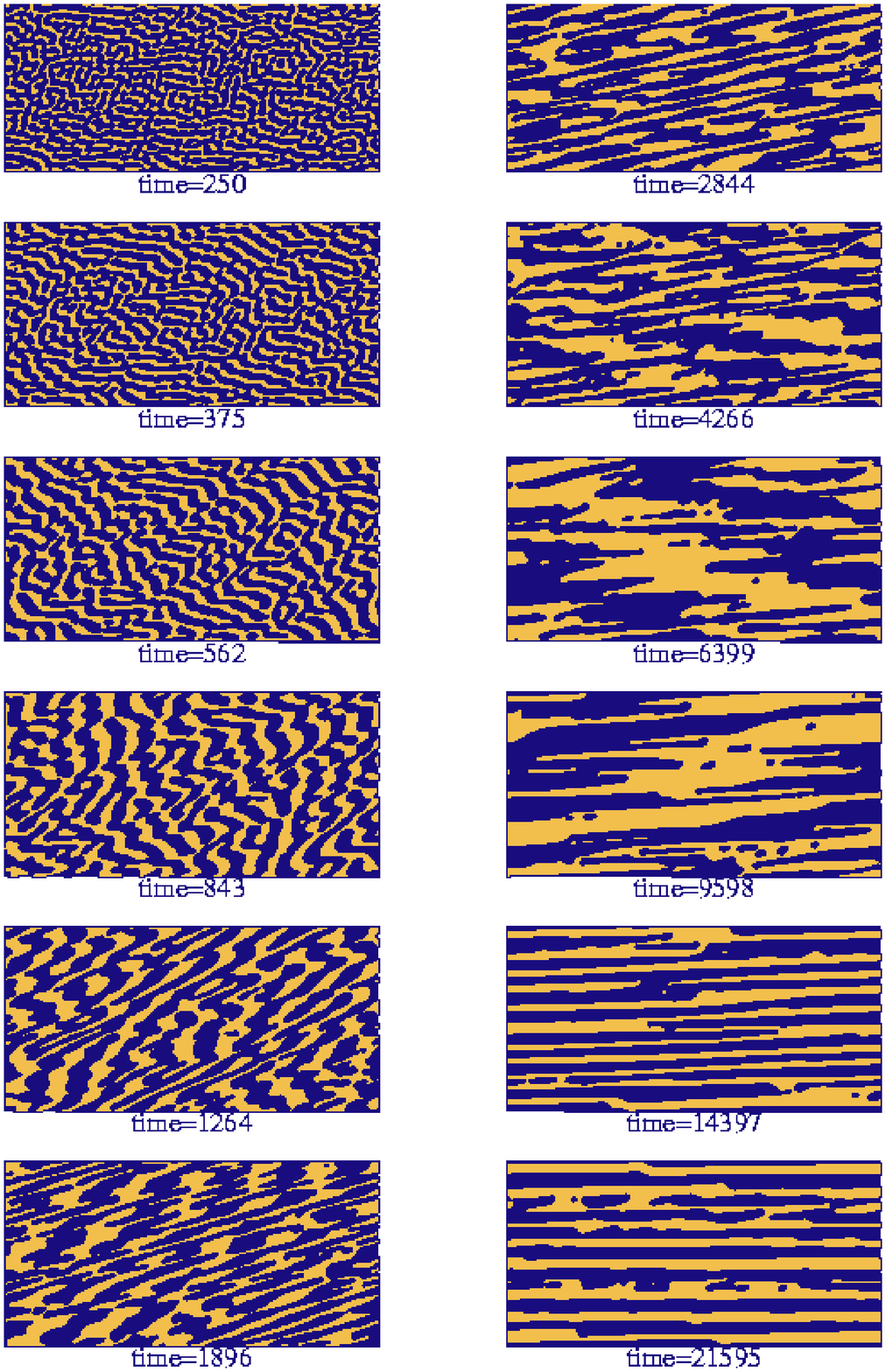,width=8cm}} 
  \begin{center} (a)
  \end{center}
\end{minipage}
\begin{minipage}[c]{9.5cm}
  \begin{minipage}[c]{4cm}
    \begin{minipage}{4cm}
      \begin{minipage}[c]{0.5cm}
        \small
        $\mbox{}\;\;\theta^*$\\  $\mbox{}\;\;\theta^\circ$\\
        \vspace{2.5cm}
      \end{minipage}\hspace{-0.3cm}\nolinebreak \hspace{0cm}
      \begin{minipage}{3.5cm}
        \centerline{\psfig{figure=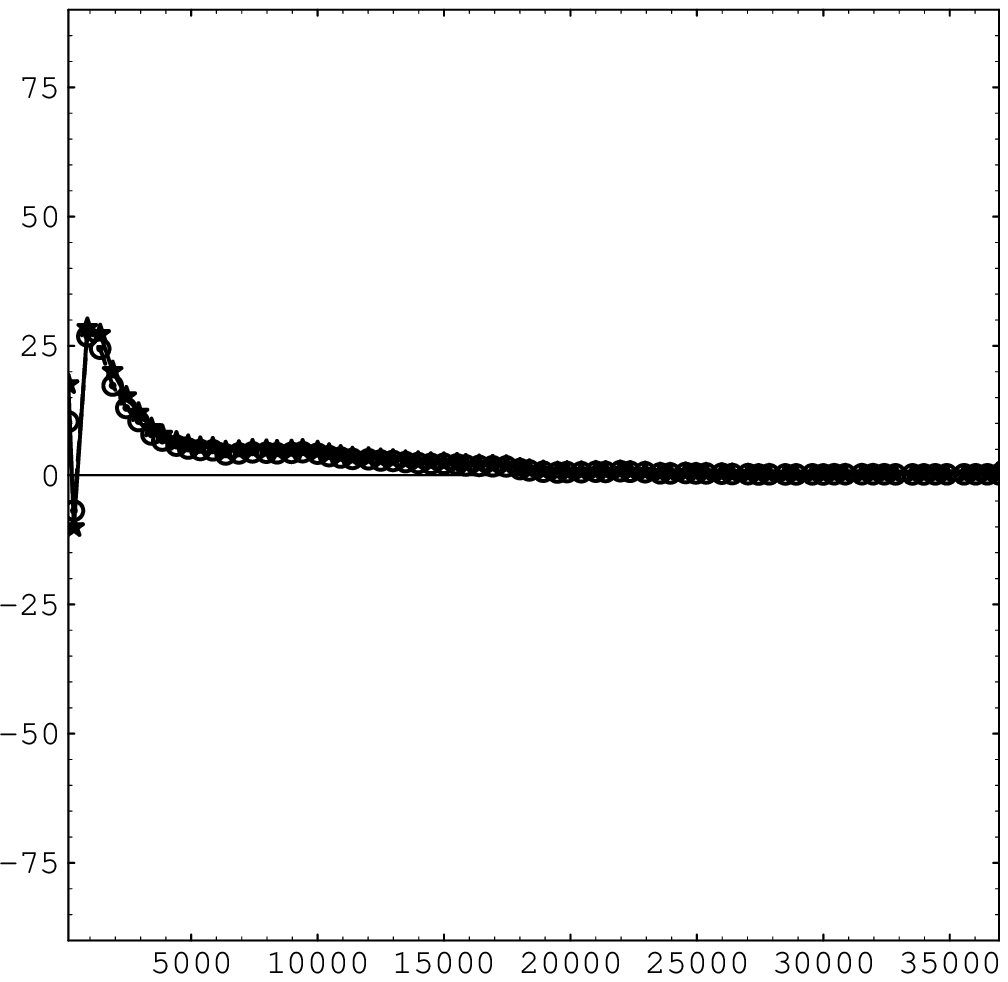,width=3.5cm}}
      \end{minipage}
    \end{minipage}
    \vspace{-0.3cm}
    \small \hfill time
    \begin{center} (b) \end{center}
  \end{minipage}
  \begin{minipage}[c]{4cm}
    \begin{minipage}{4cm}
      \begin{minipage}[c]{0.5cm}
        \small
        $R^*_{1,2}$\\$R^\circ_{1,2}$\\
        \vspace{2.5cm}
      \end{minipage}\nolinebreak \hspace{0.1cm}
      \begin{minipage}[c]{3.5cm}
        \centerline{\psfig{figure=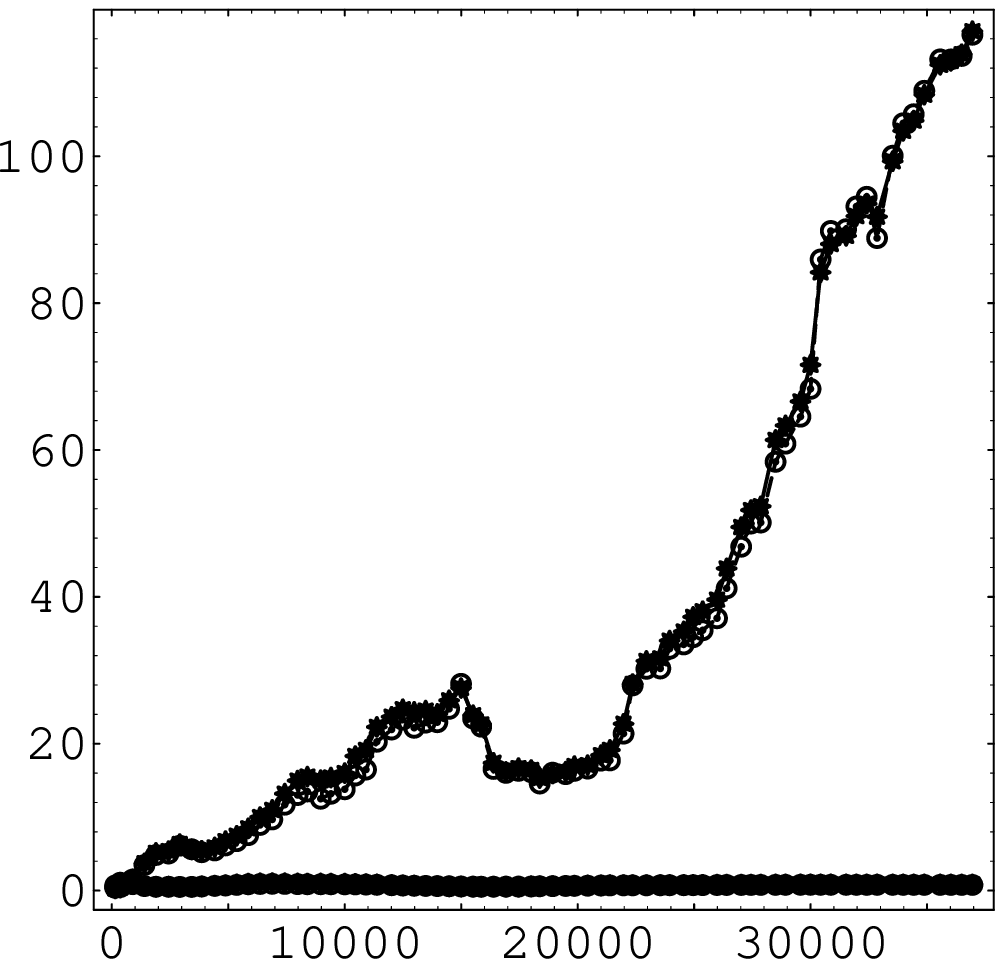,width=3.5cm}}
      \end{minipage}
    \end{minipage}
    \vspace{-0.3cm}
    \small \hfill time
    \begin{center} (c) \end{center}
  \end{minipage}
\end{minipage}\\
\caption{(a) Spinodal decomposition under shear for a high viscosity
binary fluid $(\tau_1=100,L_x=256,L_y=128)$. The high viscosity
suppresses internal hydrodynamic degrees of freedom. The shear rate is
$\dot{\gamma}=0.004$ which corresponds to a shear time $t_s=250$. (b)
Variation of the orientation (in degrees) of the pattern with time. (c) Variation
of the length scales (in arbirtary units) with time.} \label{highvisc_highshear}
\end{figure}

 The
structure develops an orientation that slowly aligns with the shear
direction while the stretching increases the length of the domains
along the shear. Once the width of the domains is smaller than the
original width of the interface the system is effectively a
homogeneous mixture.

This effect is known as shear-induced mixing. It can be observed in
the lattice Boltzmann fluids if the stretching 
effect of the shear flow is much faster
than the growth of the domains via diffusion or flow. Numerically this can be
achieved by choosing a very low mobility and a high viscosity. 
Phase separation is
suppressed because of the mixing properties of the shear flow
unless the phase-separating structure is aligned with the shear
direction. For finite lattices we sometimes observe at much later times a
nucleation of complete stripes that span the system and are
periodic in the shear direction. The time required to form these
stripes depends on the system size and it seems reasonable to assume
that this phenomenon does not occur in infinite systems.

We now consider a high viscosity fluid ($\tau_1=100$) in which
diffusive but not hydrodynamic modes are important.  The internal
dynamics leads to domain coarsening and can also prevent a complete
mixing of the system.  Figure \ref{highvisc_highshear} shows the spinodal decomposition
pattern of the high-viscosity binary mixture.  For very short times
($t<300 \sim \dot{\gamma}^{-1}$) we observe the familiar spinodal
decomposition pattern. It is, however, coarsening in a new way via
shear flow-induced collisions of the domains.  This process enhances
domains oriented in the collision direction. Then for $300<t<1000$ the
flow slowly turns the striped pattern and stretches it. At $t\sim
1000$ the rupturing of domains starts to be important and for
$1000<t<15000$ there is a continuous stretching and rupturing that
effectively stops the phase ordering process. At $t \sim 15000$ the system
developes stripes that span the system. Because periodic stripes
are unaffected by the shear flow if they are completely aligned with
it the
system now grows via the diffusion mechanism.

This evolution can be followed more quantitatively by measuring the
orientation angle and the length scales defined in Section IV. Figure
\ref{highvisc_highshear}b shows the angle of orientation to the
$x$-axis measured by $\theta^\star$ (Eqn. \ref{tetd}) and
$\theta^\circ$ (Eqn. \ref{tetl}). The two different measures for the
angle agree very well. The pattern tilts at very early times
($t<2000$) and then slowly aligns with the direction of the shear flow
as periodic stripes are created.

The graph in Figure \ref{highvisc_highshear}c  shows the
length-scales $R_{1,2}^\star$ defined in Eqns.
(\ref{rd1}) and the length scales $R_{1,2}^\circ$ defined in
Eqns. (\ref{rc1}). We very clearly see a separation of
length scales and a good agreement of the two different measures. A
minimum of the larger length scale at $t\sim 17000$ indicates the
creation of periodic stripes spanning the system. After this time
the growth of domains is no longer hindered by the continual breaking
of stretched domains.

\begin{figure}
\begin{minipage}{8cm}
  \centerline{\psfig{figure=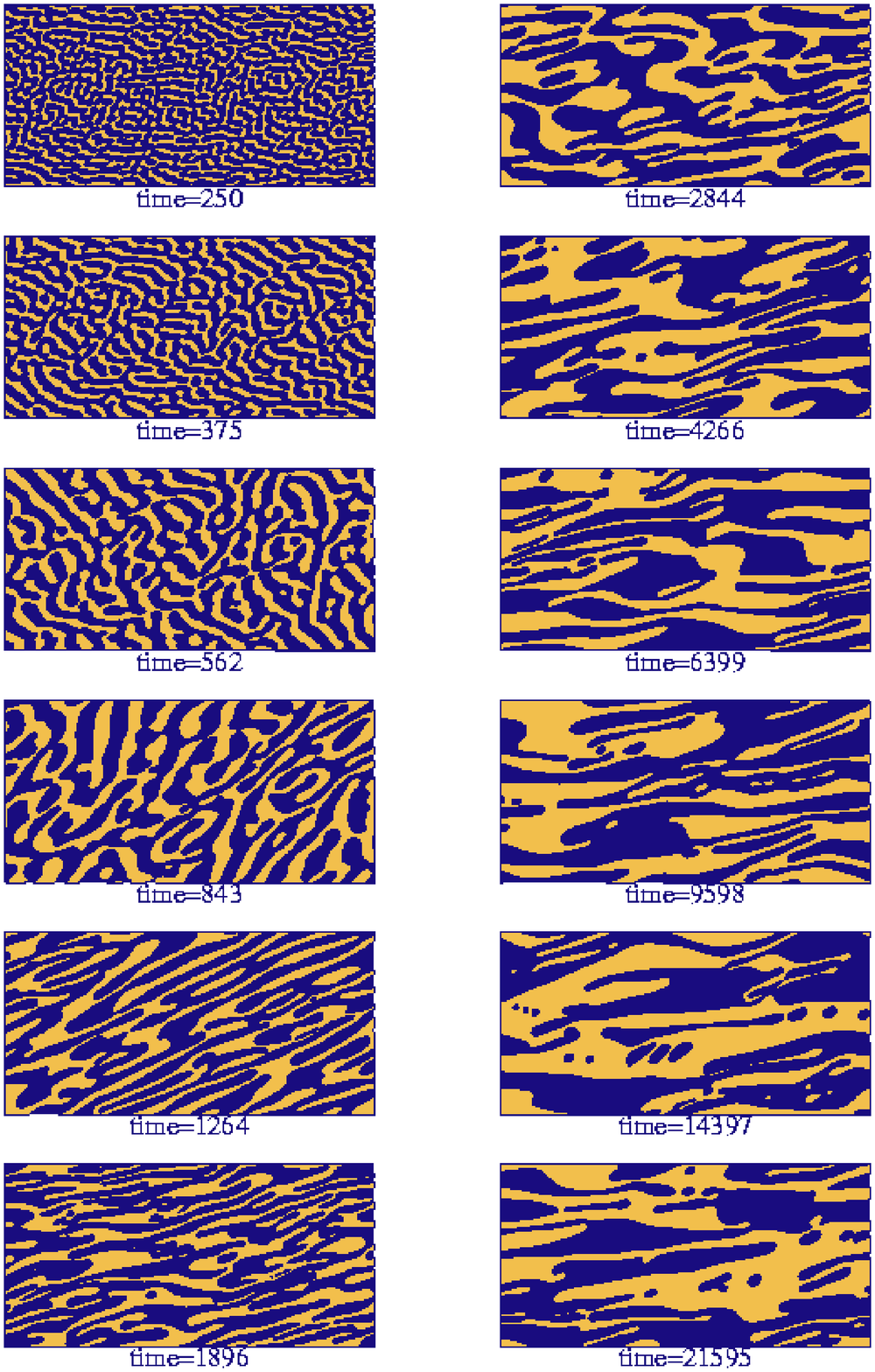,width=8cm}} 
  \begin{center} (a)
  \end{center}
\end{minipage}
\begin{minipage}[c]{9cm}
  \begin{minipage}[c]{4cm}
    \begin{minipage}{4cm}
      \begin{minipage}[c]{0.5cm}
        \small
        $\mbox{}\;\;\;\theta^*$\\ $\mbox{}\;\;\;\theta^\circ$
        \vspace{2.5cm}
      \end{minipage}\nolinebreak \hspace{0cm}
      \begin{minipage}{3.5cm}     
        \centerline{\psfig{figure=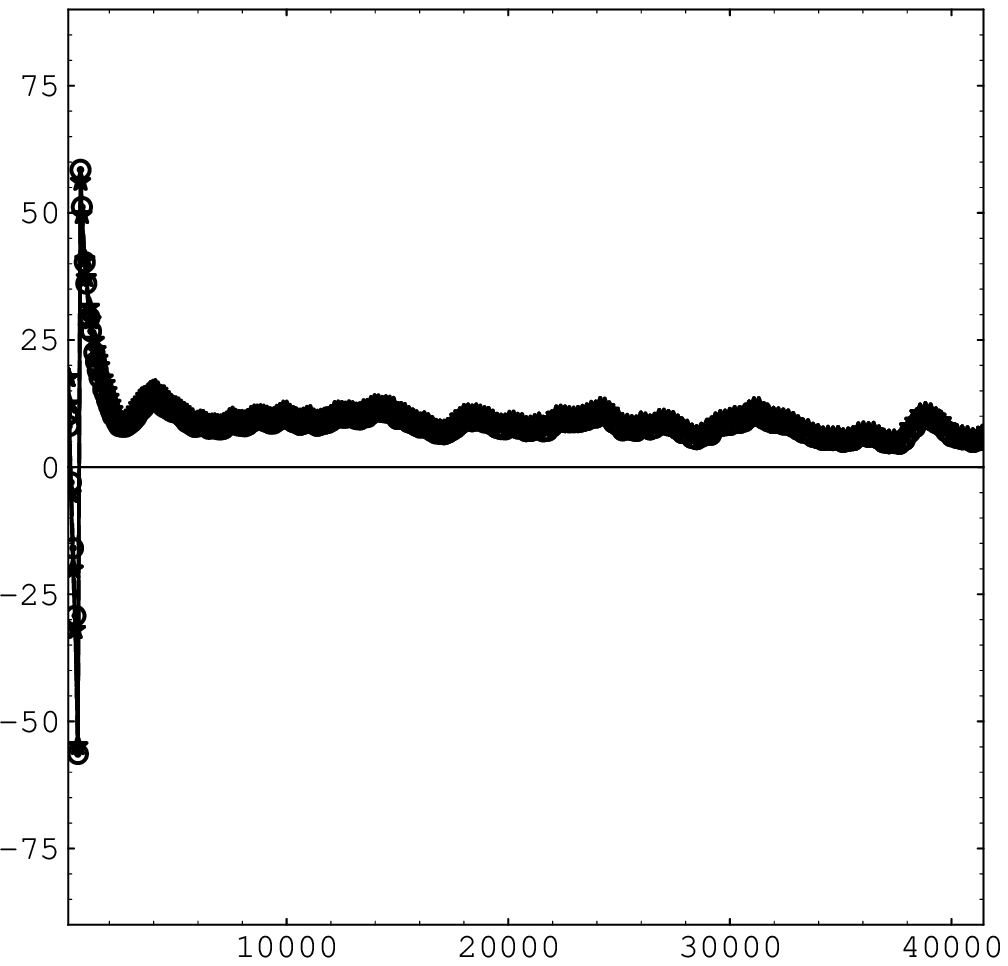,width=3.5cm}}
      \end{minipage}
    \end{minipage}
    \vspace{-0.3cm}
    \small \hfill time
    \begin{center} (b) \end{center}
  \end{minipage}
  \begin{minipage}[c]{4cm}
    \begin{minipage}{4cm}
      \begin{minipage}[c]{0.5cm}
        \small
        $R^*_{1,2}$\\$R^\circ_{1,2}$
        \vspace{2.5cm}
      \end{minipage}\nolinebreak \hspace{0.1cm}
      \begin{minipage}[c]{3.5cm}
        \centerline{\psfig{figure=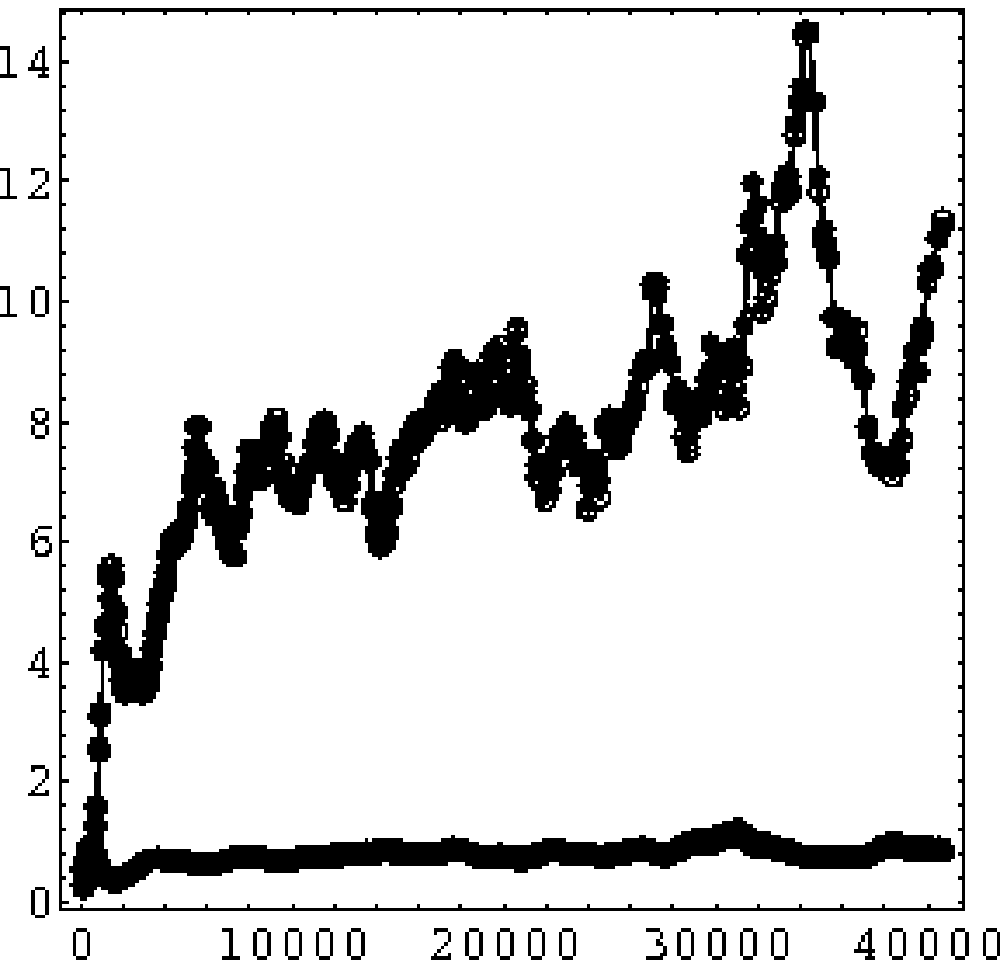,width=3.5cm}}
      \end{minipage}
    \end{minipage}
    \vspace{-0.3cm}
    \small \hfill time
    \begin{center} (c) \end{center}
  \end{minipage}
\end{minipage}\\
\caption{(a) Spinodal decomposition under shear for a medium
viscosity binary fluid $(\tau_1=1,L_x=256,L_y=128)$. The effect of internal 
flow causes the domains
to remain at an angle to the shear direction. The shear rate is
$\dot{\gamma}=0.004$ which corresponds to a shear time $t_s=250$. (b)
Variation of the orientation (in degrees) of the pattern with time. (c) Variation
of the length scales (in arbitrary units) with time.} \label{lowvisc_highshear}
\end{figure}

We now turn to consider a system with a lower viscosity that allows for a
hydrodynamic response of the domains to the shear flow. Results are
presented in Figure \ref{lowvisc_highshear}. It is
immediately obvious that the pattern differs

\noindent from that in Figure
\ref{highvisc_highshear}. The final state does not simply
consist of periodic stripes, but of dynamic structures that are constantly
stretched, broken and deformed by the flow. At least on this time
scale a state of dynamic
equilibrium is reached where the ordering effects of the spinodal
decomposition balance the disordering effects of the shear.

The quantitative measures in Figures \ref{lowvisc_highshear}b and \ref{lowvisc_highshear}c
show that after initial fluctuations the orientation of 
the pattern converges to a value that fluctuates about a
finite angle to the shear direction. This phenomenon is similar to the
behaviour of a single sheared drop that lies at a finite angle
to a shear flow\cite{stone}. 
The graph of length scales again shows a very clear
distinction between the large and small length scales. 
Strong oscillations are seen. These may be finite size effects
because the system is so
small and contains only a few domains.
However such oscillations have been seen in experiments \cite{matsuzaka} and ina model system \cite{corberi}.

\begin{figure}
\begin{minipage}{8cm}
  \centerline{\psfig{figure=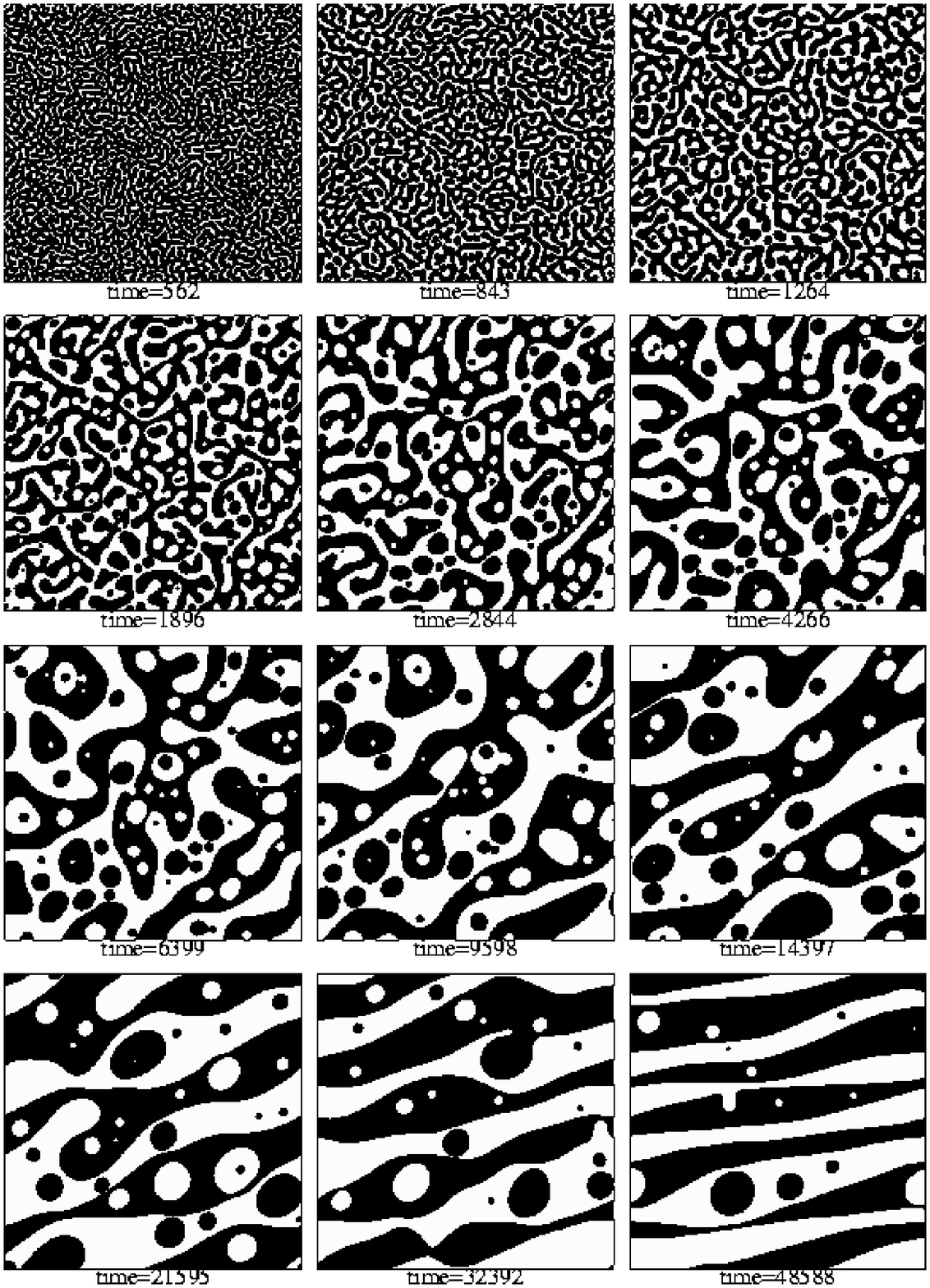,width=8cm}} 
  \begin{center} (a)
  \end{center}
\end{minipage}
\begin{minipage}[c]{9cm}
  \begin{minipage}[c]{4cm}
    \begin{minipage}{4cm}
      \begin{minipage}[c]{0.5cm}
        \small
        $\mbox{}\;\;\theta^*$\\  $\mbox{}\;\;\theta^\circ$\\
        \vspace{2.5cm}
      \end{minipage}\hspace{-0.3cm}\nolinebreak \hspace{0cm}
      \begin{minipage}{3.5cm}
        \centerline{\psfig{figure=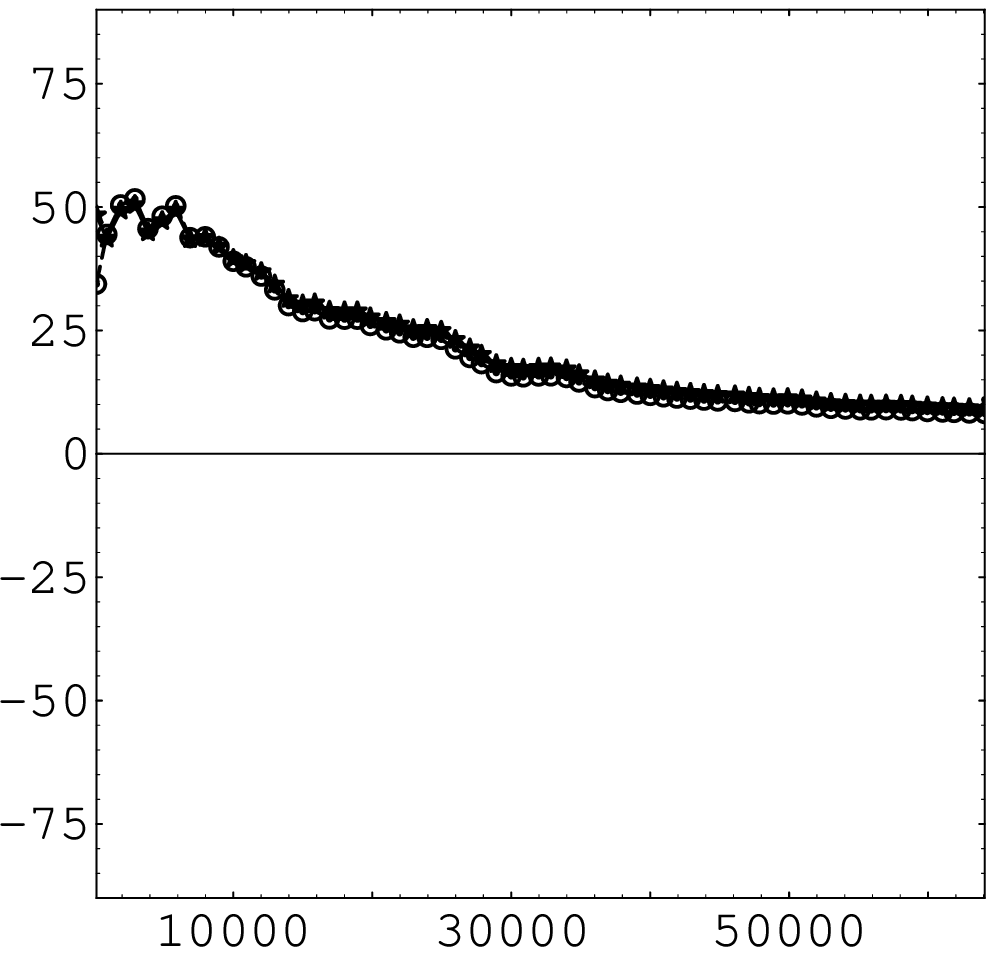,width=3.5cm}}
      \end{minipage}
    \end{minipage}
    \vspace{-0.3cm}
    \small \hfill time
    \begin{center} (b) \end{center}
  \end{minipage}
  \begin{minipage}[c]{4cm}
    \begin{minipage}{4cm}
      \begin{minipage}[c]{0.5cm}
        \small
        $R^*_{1,2}$\\$R^\circ_{1,2}$\\
        \vspace{2.5cm}
      \end{minipage}\nolinebreak \hspace{0.1cm}
      \begin{minipage}[c]{3.5cm}
        \centerline{\psfig{figure=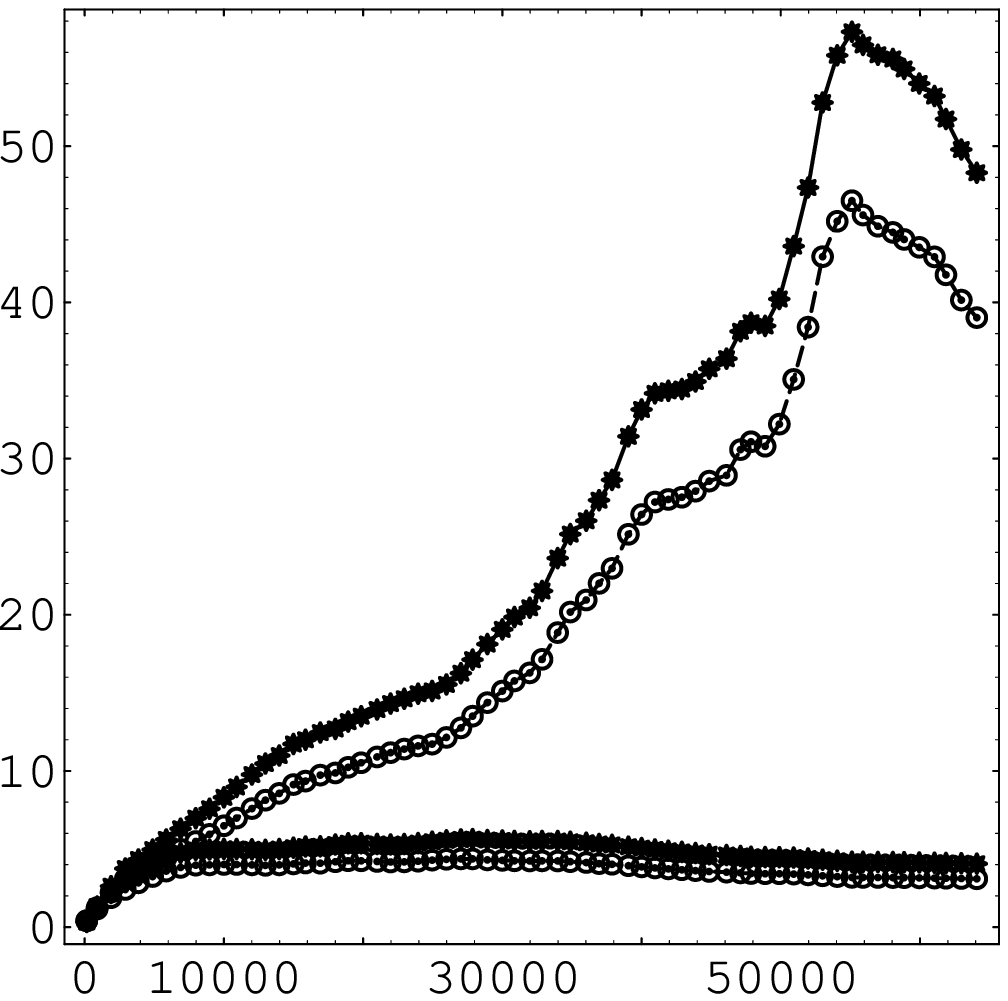,width=3.5cm}}
      \end{minipage}
    \end{minipage}
    \vspace{-0.3cm}
    \small \hfill time
    \begin{center} (c) \end{center}
  \end{minipage}
\end{minipage}\\
\caption{(a) Spinodal decomposition under shear for a high
viscosity binary fluid $(\tau_1=100,L_x=512,L_y=512)$. 
The high viscosity suppresses internal
hydrodynamic degrees of freedom. The shear rate is
$\dot{\gamma}=0.0001$ which corresponds to a shear time $t_s=1000$. (b)
Variation of the orientation (in degrees) of the pattern with time. (c) Variation
of the length scales (in arbitrary units) with time.}\label{lowvisc_lowshear}
\end{figure}

We have, so far, considered strong shear flow. Let us now consider
the same viscosity, where both diffusive and hydrodynamic flow is
possible, but lower the shear rate so that the early time
spinodal decomposition is unaffected by the flow. In
Figure \ref{lowvisc_lowshear} the spinodal decomposition for a
shear rate $\dot{\gamma}=0.0001$ is shown for a system with
$\tau_1=1$. For times
$t<1/\dot{\gamma}=10000$ we see the typical spinodal decomposition
pattern for these viscosities. 
Hydrodynamic flow leads to circular domains which then grow through
the slower diffusive mechanism. After this
time, the stretching of the domains dominates over the domain
growth and the pattern becomes non-isotropic. By $t\sim 10000$ the
pattern comprises large-stripe like domains together with the nested
pattern of drops within drops in the large domains.
As the
large domains are stretched, the drops inside them coalesce with the
walls and slowly the stripes are cleaned of the small
included drops.

These results also clearly show up in the measurements given in Figure
\ref{lowvisc_lowshear}. 
After $t>10000$ the orientation slowly
converges towards a tilting angle $\theta\sim 7^\circ$, the long and
short length scales split and the $R^\star\sim R^\circ \sim R^1 \sim
t^\frac{2}{3}$ growth law breaks down. In the $R^\#$ measure
derived from the number of domains we see a slight increase from the
normal growth law corresponding to the process of shear cleaning the
stripes from drops.

\section{Conclusions} \label{s:outlook}

In this paper we have investigated the effects of shear flow on
systems undergoing spinodal decomposition. In order to study these
systems we introduced an extension to the lattice Boltzmann 
algorithm that allows 
simulation of shear flow problems with Lees-Edwards boundary
conditions. We find that the effect of shear flow on spinodal
decomposition depends strongly on the viscosity of the fluid. Systems
with a very high viscosity tend to order in the shear direction, 
whereas systems with a lower viscosity arrive at a dynamic
stationary state where the domains lie at a finite angle to the shear
direction. 

One of the problems in simulating spinodal decomposition
under shear is that the shear flow induces long-range
correlations much faster than for un-sheared systems so
that larger lattice sizes are required to examine long-time
behaviour. 
Therefore there remain many unexplored problems concerning the structure of
spinodal decomposition under shear. For example,
it would be interesting to investigate the
transition between the sheared and non-sheared patterns for different
viscosities and to ask whether the
late-time decomposition patterns are statistically
independent of an initial shear.

\section*{Appendix A}
\label{c:pressuretensor}

We show how the full pressure tensor
(\ref{e:pressuretensor}) is derived.
The pressure of a homogeneous system is defined as the volume
derivative of the free energy. Writing the full volume dependence of
the densities $n=N/V$ and $\varphi=(N_A-N_B)/V$ explicitly we see that:
\begin{eqnarray}
P &=& -\partial_V \int_V
\psi\left(\frac{N}{V},\frac{N_A-N_B}{V}\right)
\nonumber\\
&=&-\partial_V \left( V \psi\left(\frac{N}{V},\frac{N_A-N_B}{V}\right)
\right)
\nonumber\\
&=&n \partial_n \psi +\varphi \partial_\varphi \psi - \psi.
\end{eqnarray}

For a non-homogeneous system the pressure is no longer a scalar but a
tensor. The correct form of the pressure tensor can be derived from a
Lagrangian expression for the free energy which is minimized in
equilibrium
\begin{eqnarray}
L &=& \int_V \left(\psi(n,\varphi)
 + \frac{\kappa}{2} \partial_\alpha \varphi
\partial_\alpha \varphi\right) \nonumber\\
&&+ \mu_\varphi (\int_V \varphi - (N_A-N_B)) +
\mu_n (\int_V n - N).
\end{eqnarray}
To obtain differential equations for the equilibrium we
evaluate the Euler-Lagrange equations and get
\begin{eqnarray}
\mu_\varphi &=& -\partial_\varphi \psi + \kappa \partial_\alpha
\partial_\alpha \varphi,\\
\mu_n    &=& -\partial_n \psi.
\label{chempot}
\end{eqnarray}
We multiply these equations with $\partial_\beta \varphi$ and
$\partial_\beta n$, respectively and sum the
equations. Remembering that $\mu_\varphi$ and $\mu_n$ are constants,
this yields
\begin{eqnarray}
\partial_\beta (\varphi \mu_\varphi+ n \mu_n) &=& -\partial_\alpha ( \psi
\delta_{\alpha\beta}\nonumber\\&&
 + \kappa \{\partial_\alpha \varphi \partial_\beta
\varphi - \frac{1}{2} \partial_\gamma \varphi \partial_\gamma \varphi
\delta_{\alpha\beta})\}.
\label{chempot1}
\end{eqnarray}
We then substitute the expressions for the chemical
potentials (47) and (48)
into~(\ref{chempot1})
and subtract the right-hand side from the left-hand
side to derive a tensor $\sigma$ that has a zero divergence
\begin{eqnarray}
\partial_\alpha \sigma_{\alpha\beta} &=&
\partial_\alpha ((\varphi \partial_\varphi \psi+ n \partial_n \psi -
\psi) \delta_{\alpha\beta} \nonumber\\&&
+ \kappa ( \partial_\alpha \varphi
\partial_\beta \varphi - \frac{1}{2} \partial_\gamma \varphi
\partial_\gamma \varphi \delta_{\alpha\beta} - \varphi \partial_\gamma
\partial_\gamma \varphi \delta_{\alpha\beta})).
\end{eqnarray}
For a uniform system $\sigma_{\alpha\beta} = P
\delta_{\alpha\beta}$ reduces to the homogeneous pressure. The
divergence of the pressure tensor must vanish in equilibrium. We
therefore identify $\sigma_{\alpha\beta}$ with the pressure tensor
$P_{\alpha\beta}$.

\def\jour#1#2#3#4{{#1} {\bf #2}, #3 (#4)}.
\def\tbp#1{{\em #1}, to be published}.
\def\tit#1#2#3#4#5{{#1} {\bf #2}, #3 (#4)}
\def\ap{Adv. Phys.}
\def\epl{Euro. Phys. Lett.}
\def\prl{Phys. Rev. Lett.}
\def\pr{Phys. Rev.}
\def\pra{Phys. Rev. A}
\def\prb{Phys. Rev. B}
\def\pre{Phys. Rev. E}
\def\pa{Physica A}
\def\ps{Physica Scripta}
\def\zpb{Z. Phys. B}
\def\jmpc{J. Mod. Phys. C}
\def\jpc{J. Phys. C}
\def\jpcs{J. Phys. Chem. Solids}
\def\jpco{J. Phys. Cond. Mat}
\def\jf{J. Fluids}
\def\jfm{J. Fluid Mech.}
\def\arf{Ann. Rev. Fluid Mech.}
\def\roy{Proc. Roy. Soc.}
\def\rmp{Rev. Mod. Phys.}
\def\jsp{J. Stat. Phys.}
\def\pla{Phys. Lett. A}

\end{document}